\documentclass[nonacm, manuscript, screen]{acmart}

\usepackage{xspace}
\usepackage{balance}
\usepackage{soul,xcolor}
\usepackage{url}
\usepackage{hyperref}
\usepackage{listings}
\usepackage{enumitem}
\usepackage{float}
\usepackage{array}

\usepackage[font=small]{subcaption}
\usepackage[font=small]{caption}

\newcolumntype{L}[1]{>{\raggedright\let\newline\\\arraybackslash\hspace{0pt}}m{#1}}
\newcolumntype{C}[1]{>{\centering\let\newline\\\arraybackslash\hspace{0pt}}m{#1}}
\newcolumntype{R}[1]{>{\raggedleft\let\newline\\\arraybackslash\hspace{0pt}}m{#1}}

\newcommand{\sysname}{VindiCo\xspace}
\newcommand{\spyware}{SpyCon\xspace}

\AtBeginDocument{%
  \providecommand\BibTeX{{%
    \normalfont B\kern-0.5em{\scshape i\kern-0.25em b}\kern-0.8em\TeX}}}

\setcopyright{none}

\begin{document}

\title[VindiCo]{VindiCo: Privacy Safeguard Against Adaptation Based Spyware in Human-in-the-Loop IoT}

\author{Salma Elmalaki}
%\email{salma.elmalaki@uci.edu}
\affiliation{
  \institution{University of California, Irvine}
%  \streetaddress{}
%  \city{Irvine}
%  \state{California}
%  \postcode{}
}

 \author{Bo-Jhang Ho}
 \affiliation{%
  \institution{University of California, Los Angeles}
%  \streetaddress{}
%  \city{Los Angeles}
%  \state{California}
  }
  
  \author{Moustafa Alzantot}
 \affiliation{%
  \institution{University of California, Los Angeles}
%  \streetaddress{}
%  \city{Los Angeles}
%  \state{California}
  }
  
  \author{Yasser Shoukry}
 \affiliation{%
  \institution{University of California, Irvine}
%  \streetaddress{}
%  \city{Irvine}
%  \state{California}
  }

  \author{Mani Srivastava}
 \affiliation{%
  \institution{University of California, Los Angeles}
%  \streetaddress{}
%  \city{Los Angeles}
%  \state{California}
  }

\renewcommand{\shortauthors}{S. Elmalaki, et al.}

\begin{abstract}
Personalized IoT adapt their behavior based on contextual information, such as user behavior and location. Unfortunately, the fact that personalized IoT
%context-aware apps 
adapt to user context opens a side-channel that leaks private information about the user.
%Context-aware computing is a mobile computing paradigm in which applications can adapt their behavior based on contextual information, such as user behavior, location, and activity. Unfortunately, the fact that context-aware apps adapt to user context can open a side-channel that leaks private information about the user.
To that end, we start by studying the extent to which a malicious eavesdropper can monitor the actions taken by an IoT system and extract user's private information. 
%app can monitor the adaptations triggered by authentic context-aware apps and extract user's information.
In particular, we show two concrete instantiations (in the context of mobile phones and smart homes) of a new category of spyware which we refer to as \textbf{Context-Aware Adaptation Based Spyware (\spyware)}. Experimental evaluations show that the developed \spyware can predict users' daily behavior with an accuracy of $90.3\%$. 
%a concrete instantiation of a new category of spyware apps which we refer to as \textbf{Context-Aware Adaptation Based Spyware (\spyware)}. Experimental evaluations show that the developed \spyware can predict users' daily behavior with an accuracy of $90.3\%$. %information about user's daily behavior. 
The rest of this paper is devoted to introducing \textbf{\sysname \footnote{Vindico is a latin word which means defend, protect, and punish.}}, a software mechanism designed to detect and mitigate possible \spyware.  Being a new spyware with no known prior signature or behavior, traditional spyware detection that is based on code signature or app behavior are not adequate to detect \spyware. Therefore, \sysname proposes a novel information-based detection engine along with several mitigation techniques to restrain the ability of the detected \spyware to extract private information. By having a general detection and mitigation engines, \sysname is agnostic to the inference algorithm used by \spyware.
%\sysname is implemented as part of the Android system layer and evaluated using several case studies. 
Our results show that \sysname reduces the ability of \spyware to infer user context from $90.3\%$ to the baseline accuracy (accuracy based on random guesses) with negligible execution overhead\footnote{Part of this work is published in~\cite{elmalaki2019spycon, elmalaki2018internet}}.%to $43\%$ 
\end{abstract}

%%
%% The code below is generated by the tool at http://dl.acm.org/ccs.cfm.
%% Please copy and paste the code instead of the example below.
%%
% \begin{CCSXML}
% <ccs2012>
%  <concept>
%   <concept_id>10010520.10010553.10010562</concept_id>
%   <concept_desc>Computer systems organization~Embedded systems</concept_desc>
%   <concept_significance>500</concept_significance>
%  </concept>
%  <concept>
%   <concept_id>10010520.10010575.10010755</concept_id>
%   <concept_desc>Computer systems organization~Redundancy</concept_desc>
%   <concept_significance>300</concept_significance>
%  </concept>
%  <concept>
%   <concept_id>10010520.10010553.10010554</concept_id>
%   <concept_desc>Computer systems organization~Robotics</concept_desc>
%   <concept_significance>100</concept_significance>
%  </concept>
%  <concept>
%   <concept_id>10003033.10003083.10003095</concept_id>
%   <concept_desc>Networks~Network reliability</concept_desc>
%   <concept_significance>100</concept_significance>
%  </concept>
% </ccs2012>
% \end{CCSXML}

% \ccsdesc[500]{Computer systems organization~Embedded systems}
% \ccsdesc[300]{Computer systems organization~Redundancy}
% \ccsdesc{Computer systems organization~Robotics}
% \ccsdesc[100]{Networks~Network reliability}

%%
%% Keywords. The author(s) should pick words that accurately describe
%% the work being presented. Separate the keywords with commas.
%\keywords{xx,xxx,xx}

%%
%% This command processes the author and affiliation and title
%% information and builds the first part of the formatted document.
\maketitle

\vspace{-0.2cm}
\section{Introduction}

Context-aware systems provide personalized services that are adaptive according to user context and surrounding environments. These pervasive systems have enabled a multitude of applications in several IoT sectors including smart cities~\cite{ma2017cityguard}, health care~\cite{larson2011accurate}, smart classrooms~\cite{elmalaki2021towards, taherisadr2021future}, and automotive systems~\cite{li2018planning,elmalaki2018sentio,ahadi2021adas}.
However, these enhanced capabilities come at the expense of privacy weaknesses~\cite{han2017pitchln, sicari2015security, sadeghi2015security,elmalaki2019spycon} and sometimes lack of fairness in context-adaptation~\cite{elmalaki2021fair}.

%Ubiquitous computing---that interacts and adapts to humans---is inevitable. These context-aware systems provide humans with adaptive and personalized services according to their own and their surrounding context. The new generation of these pervasive systems has enabled a multitude of applications in several IoT sectors including smart cities, healthcare, and automotive systems~\cite{ma2017cityguard,larson2011accurate,li2018planning}.
%However, this enhanced capabilities comes at the expense of more privacy weaknesses~\cite{han2017pitchln, sicari2015security, sadeghi2015security}. 

\begin{figure*}[!t]
\centering
	\includegraphics[trim={0 14.5cm 4.5cm 0}, clip, scale=0.4]{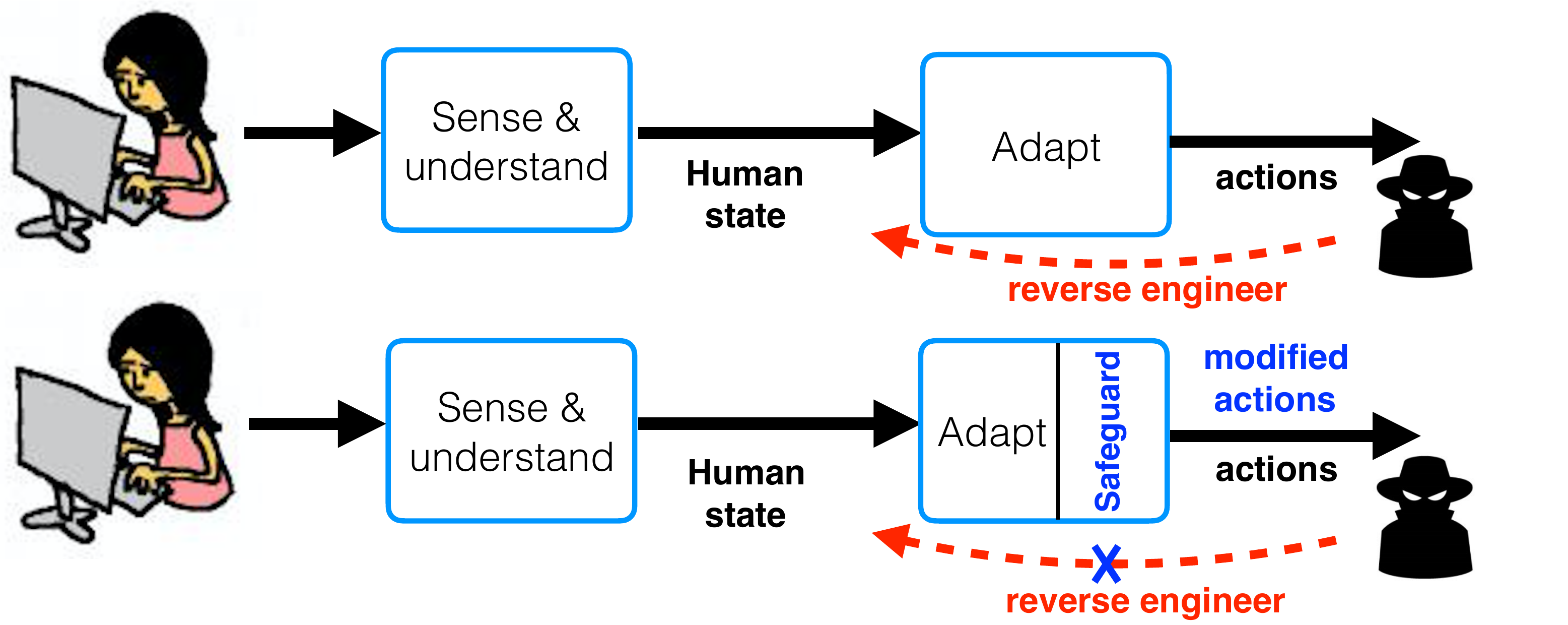}
	\caption{A cartoon that illustrates the flow of information in a personalized IoT. The IoT uses devices (typically on the edge) in order to sense and infer the human state. The IoT then adapts to the human state by applying some actions. (top) the proposed personalized IoT Spyware monitors the actions taken by the personalized IoT and reverse engineers it in order to estimate the human state (bottom) the proposed VindiCo framework which prevents modifies the IoT actions in a fashion that prevents any spyware from estimating the human state.
\label{fig:intro}}% \vspace{-7mm}}
\end{figure*}

Unfortunately, previous work ignores the fact that decisions taken by IoT are often triggered by human beings. For example, a smart NEST thermostat\footnote{NEST - https://nest.com/thermostats/} can automatically turn on and off the HVAC equipment based on users' presence. Such a coupling between human behaviors and decisions taken by IoT system can open a side channel, leaking sensitive information about users. As pictorially illustrated in Figure~\ref{fig:intro}, a human-in-the-loop (HITL) IoT utilizes edge devices (e.g., mobile phones and wearables) to sense and infer human states. Such states are then used by the IoT to produce actions and adapt its behavior to match the human state. Resting on this observation, this paper raises the following questions: (1) Can an eavesdropper who monitors the actions taken by the IoT be able to reverse engineer these actions in order to estimate human states and leak sensitive information? (2) Can we develop new software mechanisms that can detect and mitigate such privacy leaks?

To explore the answers, we introduce a new type of spyware that exploits privacy leaks in context-aware adaptations which we call \textbf{\spyware}. Besides, we purpose \textbf{\sysname}, a novel detection and mitigation engine that is designed to protect against privacy leaks due to the actions taken by HILT IoT but observed by \spyware. Contrary to traditional malware detection approaches, \sysname does not rely on prior information about the code signatures or run-time behaviors of known malware. Instead, \sysname is designed to be generic and agnostic to the implementation of \spyware.

\vspace{-0.2cm}
\subsection{Related Work}\label{sec:relatedwork}
%\r{The rapid growth and wide adoption of mobile devices have stimulated the emergence of mobile malware during the past few years. Android phones especially, due to their growing popularity, have been a target of malicious applications that invade user privacy through leaking her private data, which includes, but not limit to, location, contacts, photos, and any sort of personal identifier.}
%or monetization through making short messages or calls to premium numbers.
%\subsubsection{Kinds of Malware} 
\textbf{Context Monitoring Malware on Mobile Platforms:} Mobile systems are becoming an integral component in multiple IoT systems due to their sensing capabilities~\cite{nahrstedt2016internet}.
While mobile users benefit from these sensing technologies, there are increasing privacy and security concerns. The permission systems on both Android and iOS become the first line of defense to protect users from leaking sensitive information. However, the traditional grant-all-or-none policy allows third-party apps to have all permissions \cite{hornyack2011these}. Even worse, most users have trouble with realizing the potential privacy hazards after granting such permissions. For example, %given smartphones become a personal device, 
though seems innocuous, \texttt{ACCESS\_WIFI\_STATE} becomes a heavily privacy intrusive permission since local MAC address can serve as a unique device identifier \cite{achara2014short}. Felt et al.\cite{felt2012android} shows that as little as 17\% of users pay attention to the permissions during app installation phase.

Different side-channel attacks have been proposed, %and created many infamous activity inferencing apps.
 for example, using inertial sensors and touch screen to infer user input such as passwords \cite{han2012accomplice, miluzzo2012tapprints, owusu2012accessory, marquardt2011sp}. Besides, we witnessed how to exploit cellular signal strengths, air pressure, or power consumption for locations \cite{yan2015powerspy, ho2015pressure, michalevsky2015powerspy}, gyroscope for eavesdropping conversations \cite{michalevsky2014gyrophone}, system-level aggregate statistics for user's real world identity \cite{zhou2013identity}, and the state of shared memory for foreground apps, and even, \texttt{activity} transition sequences \cite{yuhong2015uipicker}. There is a trend that malicious apps are adapting to wearable devices \cite{raij2011privacy}. For example, MoLe \cite{wang2015mole} exploits the wrist motion derived from smartwatches to infer keystroke inputs. So far we've provided many examples showing ``Your apps are watching you'' \cite{kane2010your} in a broad spectrum which a majority of users will never realize, and for sure ``These aren't the droid you're looking for'' \cite{hornyack2011these}.

Contrary to the aforementioned side-channel attacks, we consider a spyware \emph{which does not have access to sensor information} like inertial or gyroscope sensors, a spyware which can monitor only the actions that are triggered---by HITL IoT---based on changes in these sensory data. %As we will show in Section~\ref{sec:spycon}, monitoring such actions does not always require OS permissions. 
\textbf{Malware Detection Techniques:}
%Felt et al.\cite{felt2011survey} study the intention of an app being malicious and show more than 61\% of malware try to collect user information. %personal identifiers or SMS.
 %Therefore, several techniques have been proposed to track user information flows in devices and can be broadly categorized into two groups
 Several techniques have been proposed for malware detection and can be broadly categorized into two groups. (1)\emph{Code signature-based approach} \cite{egele2011pios, grace2012riskranker, enck2011study, lu2012chex, arzt2014flowdroid} detects stealthy behavior based on the code flow. % and \r{spans anywhere between a single platform \emph{component} and considering entire system as a whole (it's not clear for me the differences of two ends)}. 
(2)\emph{Behavior-based approach} \cite{yan2012droidscope, zhu2009privacy, kang2011dta++, enck2014taintdroid} performs information leakage detection in execution time, but tackling the issue from different layers of an operating system. DroidRanger \cite{zhou2012hey} points out that an app can download binary payload in the runtime, which code-signature based approach can never diagnose its intention but raise a warning of potential hazard. 
%
%Nevertheless, both approaches can be bypassed by side channel attacks \cite{sarwar2013effectiveness} when several apps cooperatively decipher sensitive information together \cite{feng2015linkdroid}. 
Nevertheless, \sysname is distinct from the above techniques by considering the fact that a malicious app can learn sensitive information based on the adaptations made by HITL IoT.

\textbf{Malware Mitigation Techniques:}	
Different mitigation techniques have been proposed, including sensory value perturbation \cite{beresford2011mockdroid, hornyack2011these, chakraborty2014ipshield}, finer-grained permission control \cite{jeon2012drandroid, nauman2010apex}, and permission recommendation systems \cite{agarwal2013protectmyprivacy}. $\pi$Box \cite{Lee2013pibox} and SemaDroid \cite{xu2015semadroid} introduce a notion of \emph{privacy budget} and seek a balance between utilization and privacy sacrifice. \sysname shares the similar goal to quantify the degree of information being leaked and choose an appropriate data perturbation method and according mitigation magnitude based on the desired degree of data distortion.

\vspace{-0.4cm}
\subsection{Paper Contribution}\label{sec:papercont}
%The contributions of the paper are multifold and can be divided into two categories:~\\
Our contributions can be divided into two categories:~\\
%\begin{itemize}
%\setlength\itemsep{0.5em}
%\item 
\textbf{\spyware:} a new category of spyware targeting HITL IoT systems: %apps: 
	\begin{itemize}[noitemsep, nolistsep]
	\item We exploit a new side-channel attack vector arising from monitoring actions and decisions taken by IoT systems. We call this new set of attacks a \emph{context-aware adaptation based spyware}, or in short, \spyware.
	%We exploit a new side-channel attack vector arising from monitoring changes of phone adaptations by context-aware applications. We call this new set of attacks a \emph{context-aware adaptation based spyware}, or in short, \spyware. 
	\item We show two concrete instantiation of \spyware. The first instantiation targets mobile phones in which the \spyware can maliciously infer user's behavior by monitoring the decisions taken by context-based apps. We assess the performance of the developed \spyware through a one-month user study involving human subjects. The second instantiation targets smart homes, in which the \spyware monitors HVAC activity reported to the cloud and use it to infer human activity. We assess the performance of the developed \spyware through an industrial-level simulation engines simulating HVAC systems.
	%We show a concrete instantiation of a \spyware which can maliciously infer user's behavior by monitoring context-based adaptions. We assess the performance of the developed \spyware through a one-month user study. 
	\end{itemize}
\textbf{\sysname:} A safeguard against \spyware:
	\begin{itemize}[noitemsep, nolistsep]
	\item We design, implement, and demonstrate \sysname, a safeguard against \spyware. %\sysname servers to protect context-aware application from possibly leaking information spied by a \spyware. 
	We introduce a novel detection mechanism (named information-based detection) along with two genres of mitigation policies. %and discuss how \sysname decides mitigation policy and mitigation parameters. 

	\item We re-examine the proposed mitigation policies against the developed \spyware to assess it and show how its performance decreases after applying mitigation policies. 
	
%		\item We evaluate \sysname through extending Android Open Source Project (AOSP)\cite{aosp} with a new layer that supports the purposed detection mechanism and mitigation policies.
	\end{itemize}	

\vspace{-0.2cm}
\section{\spyware: A Context-aware Adaptation based Spyware}\label{sec:spycon}

\begin{figure}[!t]
\centering
  \begin{subfigure}[b]{.4\columnwidth}
    \centering
    \includegraphics[width=1\textwidth, trim={0cm 18.1cm 23cm 0},clip]{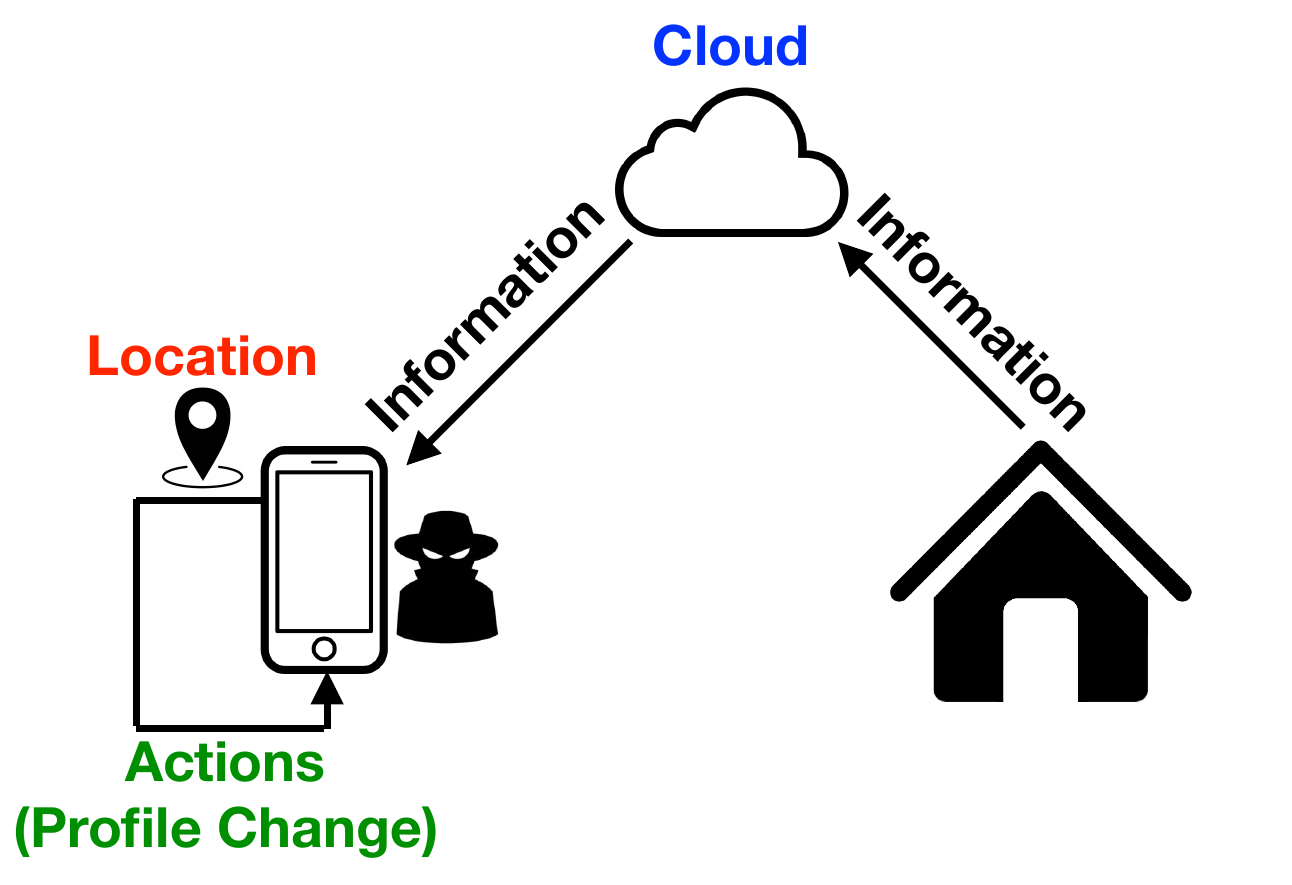}
    \caption{\small SpyCon at the Edge}
    \label{fig:SpyConEdge}
  \end{subfigure}
  \begin{subfigure}[b]{.4\columnwidth}
    \centering
    \includegraphics[width=1\textwidth, trim={0cm 18.5cm 23cm 0},clip]{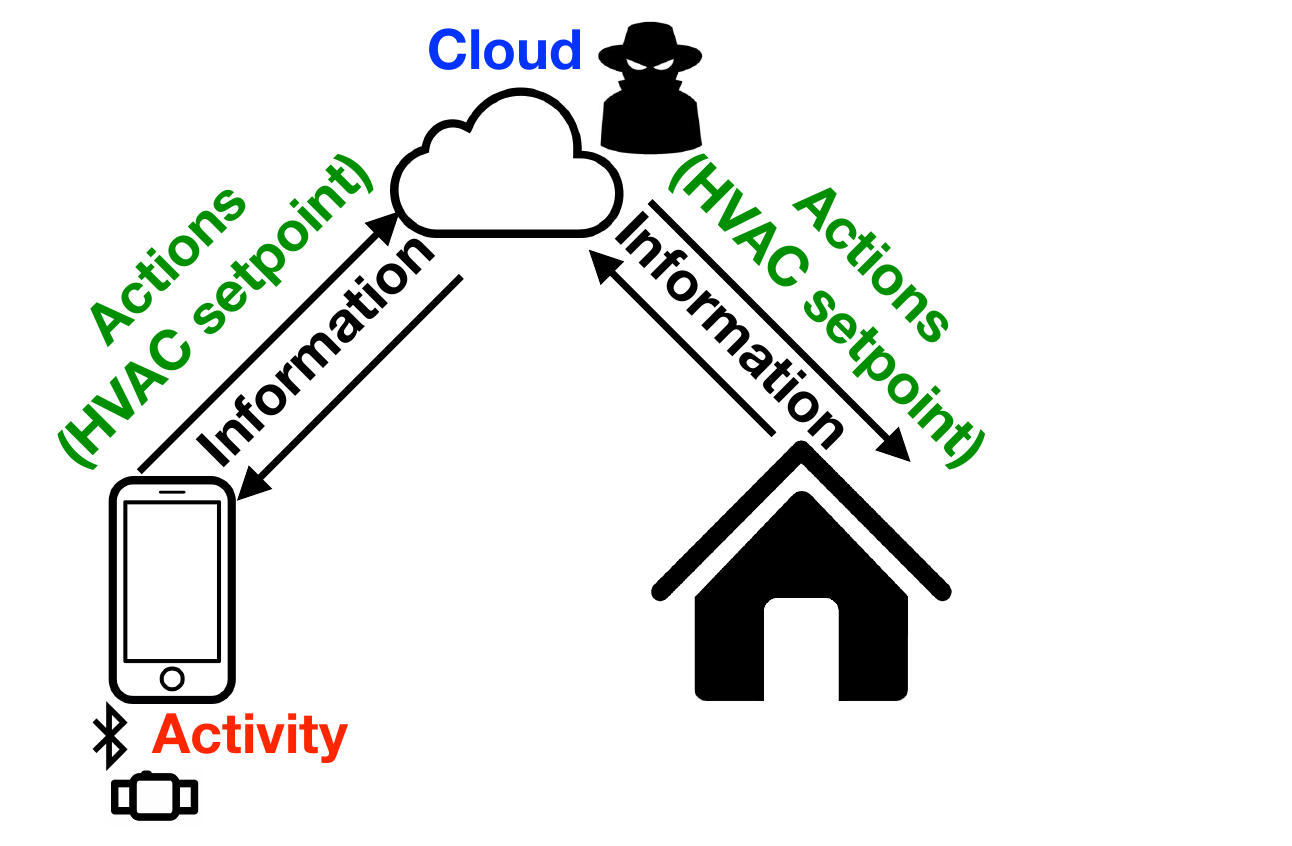}
    \caption{\small SpyCon at the Cloud}
    \label{fig:SpyConCloud}
  \end{subfigure}
  \caption{SpyCon mounted at different places (Edge, Cloud) in HITL IoT%\vspace{-0.75cm}
  }
  \label{fig:SpyConEdgeCloud}
\end{figure}

%This section serves as a detailed explanation for an example of a \emph{Context-Aware Adaptation Based Spyware (\spyware)} that is designed, implemented and studied to highlight the possibility of \spyware on context-aware applications. (We will visit this example again in our evaluation in Section~\ref{sec:evaluation} to show the effect of out mitigation techniques). The objective of the designed \spyware is to infer user's behavior by monitoring adaptations triggered by context-aware applications.

%This section introduces the Context-aware Adaptation based Spyware (\spyware) targeting HITL IoT and illustrates how it works and its potential adverse outcomes. Thanks to the increase in sensing capabilities of new mobile phones, they are heavily used as a rich sensory platform for multiple HITL IoT. To this end, we show two examples of \spyware, one \spyware example that targets IoT mobile/edge devices while the other one targets the IoT cloud. For the remainder of the paper, we refer to these two examples of \spyware as ``SpyCon Edge'' and ``SpyCon Cloud''. % app that stealthily learns 

%the semantics of the user locations inferred by those adaptions made by other context-aware apps. %Though we use location as an example, the same concept can be generalized and applied to collecting other types of sensitive user data.

We define a spyware that \emph{spies} on context-aware adaptations a \textbf{\spyware}. Since adaptions are dependent on the user context, observing adaptions made by HITL IoT can reveal sensitive user information. In this section, we show two examples of \spyware, one targets IoT mobile/edge devices while the other one targets the IoT cloud~\cite{elmalaki2019spycon}. We refer to these two \spyware as ``SpyCon Edge'' and ``SpyCon Cloud'', respectively.

% To better understand the scale of such malware, we provide an analysis on the context-aware applications in the mobile software markets. 

%We will give in this section an example of ConMan, as a motivation example, that changes the phone settings based on the user's location. This example is motivated by the huge amount of software in the market that performs phone settings adaptation as shown in Section ~\ref{}. 

%\subsection{ConMan Malware}
%ConMan is based on monitoring the set of actions taken by a context-aware application in order to infer information about the user without his permission/knowledge. 
\vspace{-0.2cm}
\subsection{\spyware Edge: Popular Phone Manager Apps}\label{sec:spyconedge}
%Mobile phones are becoming an integral component of many IoT systems by acting as a sensory rich edge device. In this example, we introduce \spyware mounted on a mobile phone. In Figure ~\ref{fig:SpyConEdge}, we show a scenario of an IoT application that incorporates information streamed from the user house along with sensory data collected from his phone to recommend actions to the IoT user. One example of such scenarios, is using the user location (collected from the phone) along with information from smart appliances at his home (smart fridge) to send the human a notification about required grocery list or adapt the WiFi settings based on the human's location. 

A typical pipeline of an IoT application includes edge devices which generate sensor data, an IoT cloud which stores and forwards these data, and a mobile phone which consumes the data, computes the actions, and presents them to IoT users.
Mobile phones are arguably a primary target for spyware due to their sensing capabilities and the integrations with IoT systems.
In this scenario, we consider \spyware is mounted on a mobile phone to leak sensitive information, depicted in Figure~\ref{fig:SpyConEdge}.
The \spyware, for example, can reveal a user location such as next to the smart fridge (the sensitive information) if a notification reminder of a grocery list is observed (the adaption).
Location-based phone settings management is one of the most popular context-aware applications\footnote{By the time this paper was written, context-aware phone settings management applications
%applications that adapt the phone's settings based on your location or your time preferences 
ranked 3rd in the Productivity category in the Android Developer Challenge~\cite{tasker}.}. Due to their capability to adapt to user context, apps like Llama~\cite{llama}, Tasker~\cite{tasker}, and Locale~\cite{locale} have gained more than 1 million downloads from Google Play Store.
Moreover, these location-based apps are increasingly integrated as part of larger IoT systems. %For example, Tasker can be combined with IBM Watson IoT platform to allow HITL IoT to take location-based decisions~\cite{taskerIoTWatson}.
%These context-aware apps provide a friendly UI for users to define their profiles. 
%A \emph{profile} contains a context-based trigger and a set of actions. Upon the phone context matches the trigger, these apps perform the corresponding actions to change the \emph{phone settings}. For instance, common profile configurations are muting the ringer volume when a user is in class or a conference room or enabling WiFi whenever the user enters home. 
Motivated by the popularity of these location-based context-aware apps, we choose user location as the sensitive data for which \spyware is trying to leak.  %In particular, we show in the next subsection, how a malware running on a phone can extract user's behavior by monitoring the changes in phone settings triggered by those location-based phone manager apps.

\begin{table}[t]
\centering
\small
\begin{tabular}{| c | c  || c | c |}
\hline
\textbf{PS} & \textbf{Description} & \textbf{PS} & \textbf{Description} \\
%\textbf{Symbol} & & \textbf{Symbol} & \\
\hline
R & Ringer mode & P & Wallpaper \\ \hline
H & Touch sound &D & Dialpad sound \\ \hline
W & Enable WiFi &  A & Alarm volume\\ \hline
I & Ringer volume & M & Media volume\\ \hline
T & Display timeout & B & Screen brightness \\ \hline
V & Vibration on touch   & L & Screen locking sound \\ \hline \hline
%, (P) home screen wallpaper, (D) enable dialpad tone, (L) enable screen locking sound, (H) touch sound, (V) vibration on touch, and (T) display timeout.
\end{tabular}
\caption{Phone settings (PS)}%\vspace{-5mm}
\label{tab:phonesettings}
\end{table}

\vspace{-0.1cm}
\subsubsection{\textbf{Spyware Description}}
 \label{sec:spywareDescription}
 We design a \spyware that monitors changes in phone settings to demonstrate it is possible to leak user location, an arguably the most sensitive type of user information~\cite{locationPrivacyAct}.
The phone setting changes are triggered by the decisions taken by a location-based context-aware app as part of an IoT system.
%We are interested in designing a \spyware that monitors changes in phone settings---which are triggered by the decisions taken by a location-based context-aware app as part of an IoT system---and uses these changes to leak user's location, arguably one of the most sensitive user information~\cite{locationPrivacyAct}. 
We start by making the following two important remarks: %\r{to describe our threat model}:~\\%\vspace{1mm} ~\\
\begin{itemize}
	\item \textbf{No user permissions:} Unlike location information, many phone settings can be monitored without seeking user permissions. For example, \spyware can easily get current screen brightness or alarm volume without user consent.%\vspace{1mm} ~\\
\item \textbf{Ambiguity on setting changes:} Manual adjustment can make changes in phone settings through physical buttons. Although \spyware can not discriminate \emph{a priori} between the changes in the phone settings done by a location change or by manual adjustment from users, machine learning algorithms can be handy in discovering repetitive patterns in the data. 
\end{itemize}
% and hence infer user's daily behavior.
%We start by noting that many phone settings can be monitored without seeking user permissions. A \spyware can easily get current screen brightness or alarm volume without user consent. As a role of a spyware, to retrieve user context, our \spyware observes changes in phone settings which may be triggered by changes in user locations.
%However, changes in phone settings can be made by other factors, such as a user manually adjusts ringer volume through physical buttons. Although a spyware can not discriminate \emph{a priori} between the settings' change that are triggered by location change and by the changes that are manually performed, machine learning algorithms can be handy in discovering repetitive patterns in the data and hence infer user's daily behavior. 

%\vline

\noindent The operation of the designed \spyware is divided into two phases:%\vspace{1mm} ~\\
\begin{itemize}
	\item \textbf{Logging:}% In this phase, %the spyware registers to all phone setting listeners. A list of all phone settings that we considered in our spyware are given in Table~\ref{tab:phonesettings}. %As noted before, this operation does not need user consent. 
%Next, 
%\spyware monitors all the changes in phone settings and records the timestamp of the change in a phone setting along with its value. 
\spyware monitors all the changes in phone settings and records a timestamped value upon a change is detected. A list of phone settings that we consider in our \spyware is given in Table~\ref{tab:phonesettings}.%\vspace{1mm}  ~\\
%\noindent\textbf{- Clustering:} Once enough data are collected, the recorded data are analyzed.  In our implementation, we use a k-means clustering algorithm to discover patterns in the collected data and hence infer user's daily behavior. 
%More details of the implemented clustering algorithm are given in Section~\ref{sec:sca} after we discuss the user study setup.
\item \textbf{Data Mining:} Once enough data is collected, \spyware analyzes these data to discover repeated patterns and hence infers user's daily behavior.
More details about the data mining algorithm are given in Section~\ref{sec:sca} after we discuss the user study setup.
\end{itemize}

\vspace{-0.1cm}
\subsubsection{\textbf{\spyware User Study}}\label{sec:userStudy}
%To study how a \spyware can invade and steal users' privacy by monitoring adaptations made by context-aware adaptation applications,
% We developed two   applications, a \b{shallow logging} app and a \spyware. % malware to breach user privacy. 
  \vspace{-0.1cm}
\paragraph{\textbf{Shadow Logging Application}} \vspace{-0.12cm}
\label{sec:auth_app}
To understand how much information is leaked by context-aware apps like Tasker and Locale, we developed a shadow app that resembles the functionality of Tasker and Locale in order to collect the ground truth data. First, we ask users to enter the same \emph{profiles} which they would provide in the context-aware apps (Tasker or Locale). To be more specific, users have to enter a fixed-radius circular geo-fence as a \emph{context} trigger, as well as a set of \emph{actions} (e.g., adjusting screen brightness or changing ringer mode to vibration) that would be activated when the user enters these geofences. The full phone settings we considered are listed in Table \ref{tab:phonesettings}. Secondly, in order to keep track of the golden output (ground truth) for later evaluation, the shadow app keeps and timestamps a record whenever the active profile is changed, implying that the user moves to a different location.

\vspace{-0.15cm}
\paragraph{\textbf{SpyCon Application}}
\label{sec:spycon_app}
%Besides the authentic context-aware adaption application, we also developed a \spyware. 
We developed a \spyware whose only task is to log phone settings in the background \emph{without any interaction} with all the other apps, including the context-aware app\footnote{In the real world, this \spyware can provide some functionality but collect data stealthily, which is a typical way a spyware hides its true intention.}. All the settings collected by the \spyware app can be accessed \emph{without permissions} in Android OS, including those listed in Table \ref{tab:phonesettings}.

However, it should be noted that any \spyware app may 
benefit from knowing whether a context-aware adaption application is installed. This information can be retrieved  through different ways such as the \texttt{getInstalledApplications()} API
\footnote{
Even though Android may protect this API by adding a permission in the future, studies have shown that it is hard for most users to relate the side-channel privacy implications to the granted permissions in different apps~\cite{felt2012android}.}.

%Second, the malicious app needs to know what context adaptations are involved in the context-aware apps. Typically, this can be fetched from the context-aware application store (e.g., ``Google Play'') page. 
%For example, Tasker~\cite{tasker} and Locale~\cite{locale} apps specify that phone settings are adapted based on user location.

\begin{figure*}[!t]
\centering
  \begin{subfigure}[b]{.32\textwidth}
    \centering
    \includegraphics[width=1\textwidth, trim={0 0 0 0},clip]{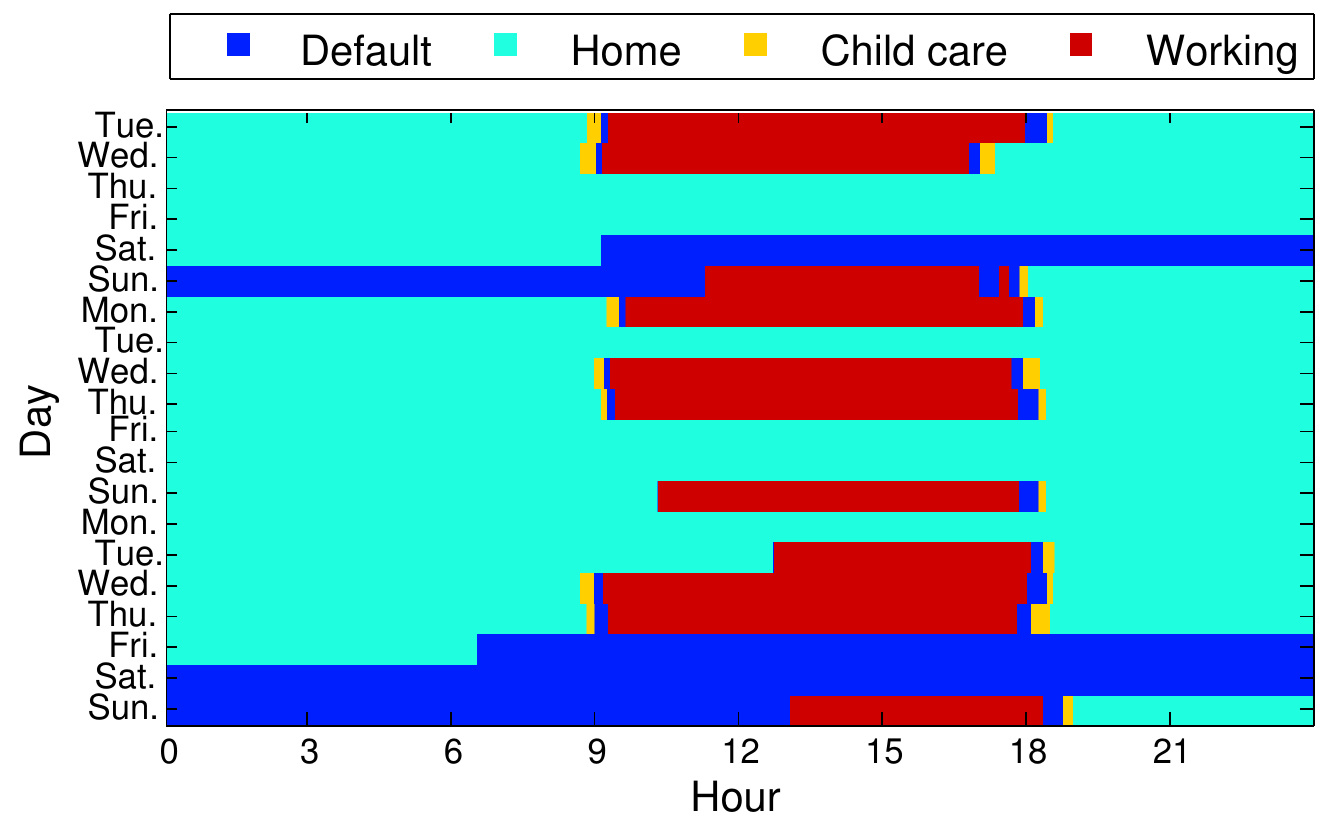}
    \caption{\small Golden output }
    \label{fig:profile_timeline_golden}
  \end{subfigure}
  \begin{subfigure}[b]{.32\textwidth}
    \centering
    \includegraphics[width=1\textwidth, trim={0 0 0 0},clip]{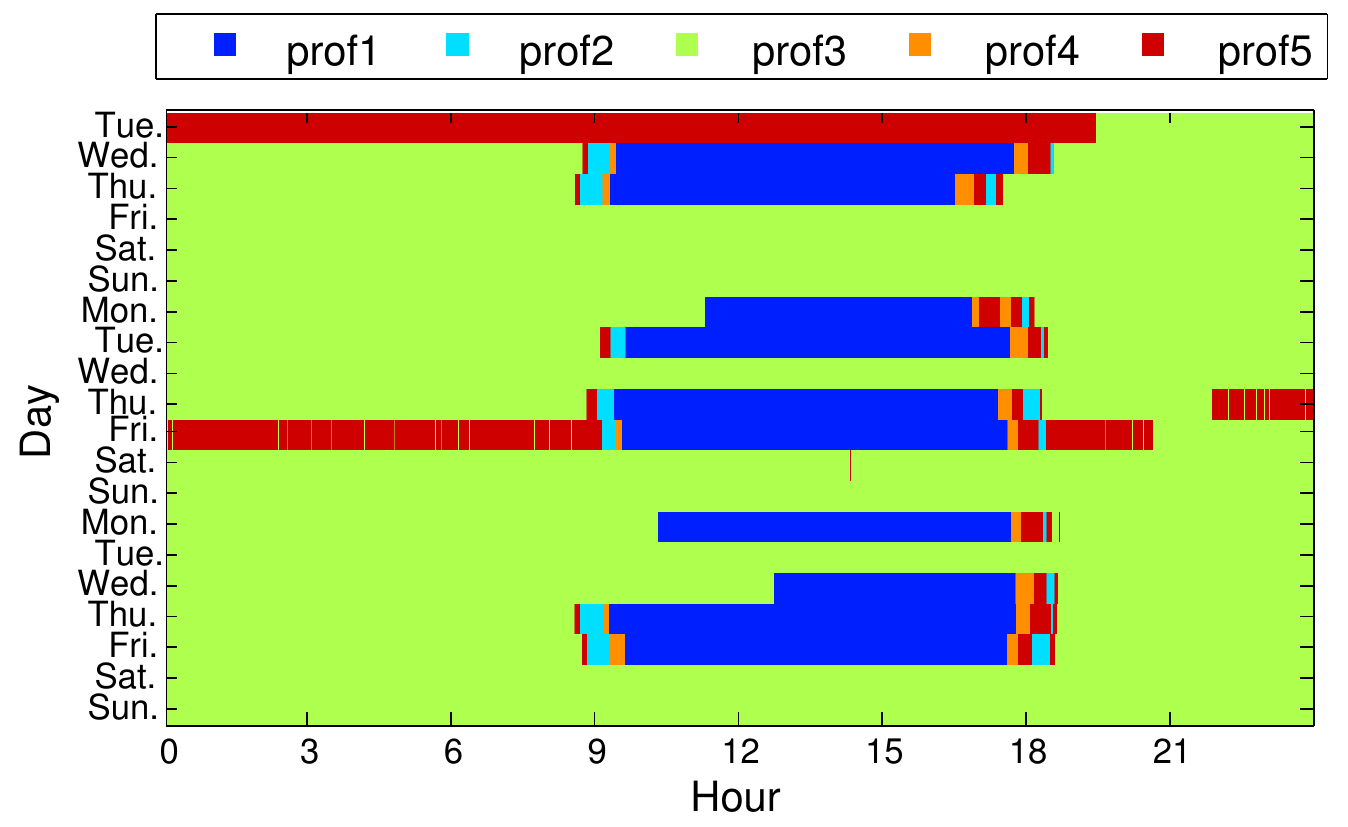}
    \caption{\small Clustered by all features }
    \label{fig:profile_timeline_full}
  \end{subfigure}
  \begin{subfigure}[b]{.32\textwidth}
    \centering
    \includegraphics[width=1\textwidth, trim={0 0 0 0},clip]{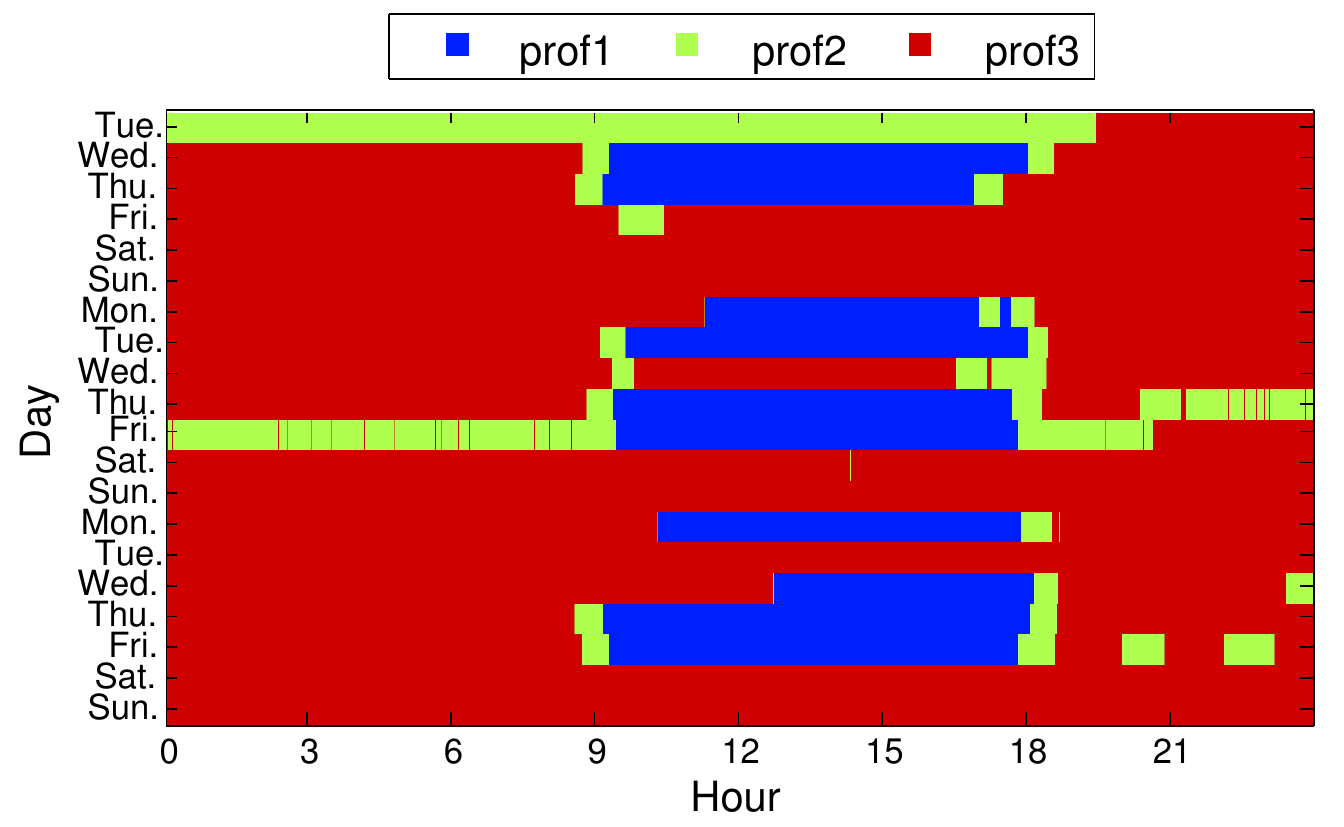}
    \caption{\small Clustered by dominant features}
    \label{fig:profile_timeline_selected}
  \end{subfigure}

\caption{One example of profile timeline from user \#2. \vspace{-0.4cm}}
\label{fig:profile_timeline_example}
\end{figure*}

\vspace{-0.15cm}
\paragraph{\textbf{User Study}}
We implemented both apps mentioned above on Android 5.0.1 running on Nexus 4 and Nexus 5. Seven participants are recruited in our user study, including four males and three females. Each participant carries our phone for four weeks. Users can choose the settings/profiles based on their personal preferences, and they are allowed to change the phone settings manually.
Based on the data we collected during the user study, we explore what sensitive information can be mined maliciously as shown in the following experiment. 

\vspace{-0.2cm}
\subsubsection{\textbf{Experiment 1a: Data Mining by Clustering}}\label{sec:sca}
%\r{Bo: do we need a figure to explain here?} 
%We implemented both aforementioned apps on Android 5.0.1 and run on Nexus 4 and Nexus 5. Seven participants are recruited in our user study, including four males and three females. Each participant is required to carry our phone for four weeks. Based on the data we collected during the user study, we perform a clustering algorithm and explore what sensitive information can mined maliciously. 
 
Revealing the semantics of the user location trace, or equivalently, the active profile sequence from phone settings is challenging since both the profile and the phone settings do not always have a one-to-one mapping. This is because (1) Users  configure only a subset of the $12$ settings listed in Table \ref{tab:phonesettings} in their profiles and hence it is not known a priori which subset of settings are used by the user. Furthermore, different profiles may include different phone settings to be changed. (2) Users can manually override the phone settings by, for instance, pressing the volume buttons or adjusting the brightness through the \emph{Settings App}.
Thus, we use a clustering technique to approach the user data mining problem, and in particular, we use k-means algorithm.

% (1) users can manually change some phone settings regardless of their location, (2) users usually set only a subset of the $12$ settings (listed in Table \ref{tab:phonesettings}), and (3) two different profiles may update different sets of phone settings, which means some settings from the first action may still exist after the second action is applied. %Nevertheless, we hypothesize that the phone settings updated by the same profile are similar.

%There are several challenges to apply k-means. First, we need to turn a phone setting into a vector. For phone settings which are continuous (i.e., can be represented by a real number), we normalize the number between 0 to 1. For settings which are discrete such as ring tone, there is no order among possible values and cannot be normalized. To address issue, we expand the value into a multi-dimension vector, each dimension refers to a possible discrete value. The conversion is being done by setting the dimension whose associated value is the same with presenting value and putting zeros to all the other dimensions. This guarantees that two different values will have the same euclidean distance in the vector space.

Deciding the number of clusters in the k-means algorithm is known to be hard in general and is usually application dependent. %To study this, we collected the golden output from our users. The \emph{golden output} is packed as records; each record contains a time span and a profile. By analyzing the profiles of all users, we conclude that three profiles are common among all the participants, which correspond to home, working, and default (which is usually in transportation). Given that these three profiles span vast majority of time among all users, a good $k$ is either 2 or 3.
 Since our \spyware does not know how many profiles are defined by users, we  brute-forcedly set $k$ to be any value between 2 through 7 (selected based on the maximum number of profiles defined by our participants). Our algorithm returns the clustering result with the highest silhouette score.

\vspace{-0.1cm}
\paragraph{\textbf{Critical Phone Settings}}
%We are also curious that, from a \spyware point of view, whether monitoring all phone settings is necessary or not. A phone setting may reveal more information than another. 
 Inspired by how most unsupervised machine learning algorithms work, we implement a greedy algorithm to find dominant phone settings. %since some phone settings may reveal more information than others. 
 The algorithm procedures are provided below: 
\begin{enumerate}[noitemsep, nolistsep]
   \item Initialize the selected feature set $S=\phi$.
   \item We examine every other setting $f$ not in $S$ by performing k-means with feature set $S \cup \{f\}$. The silhouette score $h_f$ is computed accordingly.
   \item Denote $\hat{h}$ as the maximum $h_f$ from the previous step. If $\hat{h}$ is larger than previous silhouette score, then $S = S \cup \{f\}$ and go back to step 2. Otherwise, the algorithm terminates.
\end{enumerate}

\begin{table}[!t]
\small
\centering
\begin{tabular}{| c || l | l | l | l || l |}
\hline 
 UID& \multicolumn{4}{c||}{\# clusters using all features } & \multicolumn{1}{c|}{Dominant} \\
% & \multicolumn{4}{c||}{ } & \multicolumn{2}{c|}{features} \\
   \cline{2-5}
  & base & 2 & 3 & 2-7 &  \multicolumn{1}{c|}{features}  \\ \hline \hline
1 & 75.2 & +18.9 & +22.9 & +19.1 & +21.8  W,R,V \\ \hline
2 & 56   & +17.2 & +24   & +18.3 & +24.1  R,B,W \\ \hline
3 & 80.5 & +12.9 & +14.4 & +13.6 & +16.7  R \\ \hline
4 & 45.6 & +37.3 & +34.2 & +35.6 & +35.9  W,R,L\\ \hline
5 & 42   & +24   & +35.2 & +24   & +41.8  T,R,A \\ \hline
6 & 57.9 & +4.4  & +36   & +4.4  & +40.7  A,R,B,W\\ \hline
7 & 78   & +15   & +15.5 & +15   & +15.6  R,O\\ \hline
\hline
\textbf{Avg.} & \textbf{62.2} & \textbf{+18.5} & \textbf{+25.8} & \textbf{+18.2} & \textbf{+28.1} \\ \hline
\end{tabular}
\caption{Clustering accuracy (in percentage) of all users compared to the baseline accuracy (the accuracy that the \spyware can have based on blind guesses) by applying k-means using the settings in Table~\ref{tab:phonesettings}\vspace{-6mm}}
\label{tbl:kmeans_accuracy}
\end{table}

\vspace{-0.1cm}
\paragraph{\textbf{Privacy Implications}}\label{sec:privacyImplication}
 The clustering result of one participant in our study is demonstrated in Figure \ref{fig:profile_timeline_example}. Figure \ref{fig:profile_timeline_golden} shows the actual user profile changes across the day (the golden output as explained in Section~\ref{sec:auth_app}). Figure \ref{fig:profile_timeline_full} shows the k-means clustering result (using an adaptive number of clusters) and demonstrates similar patterns with the golden output in Figure~\ref{fig:profile_timeline_golden}. %Note in this result we use adaptive number of clusters.
  Our algorithm is able to capture subtle events, for example, learning that the user regularly went to a particular place (which turns out to be the child care) after leaving or before returning home, despite the portion of time this user spent in child care is short. Clustering result derived by dominant features from our feature selection algorithm is shown in Figure \ref{fig:profile_timeline_selected}. %Generally speaking, there is no big difference compared with clustering with the full feature set. However, the clustering result becomes noisier and less number of clusters are chosen, indicating the system sacrifices interesting facts to enhance accuracy. 
Figures \ref{fig:profile_timeline_full} and \ref{fig:profile_timeline_selected} clearly indicate the ability of the developed \spyware to reconstruct user context (switching profiles in this case) by just monitoring its side effect (changes in phone settings)\footnote{If the user specifies two profiles with the same settings, \spyware will recognize them as the same profile. However, the incentive of the user to define the same settings for multiple profiles defies the idea of the context-aware app.}.

 The overall accuracy of our clustering algorithm is reported in Table~\ref{tbl:kmeans_accuracy}. 
 %Defining the accuracy is tricky since although k-means can group several time spans into the same profile, it fails on interpreting the semantic meaning of each cluster. Therefore, we enumerate all profile name assignments and compute their similarities with the golden output. The accuracy is defined as the highest similarity among all the assignment. 
We define \emph{baseline accuracy} by using blind guesses, that is, the \spyware always reports a user is at home without observing any phone settings. The results in the rest of the columns are the additional information (the increase in accuracy) the \spyware gains over the baseline accuracy if an inference is used based on a different number of clusters. The accuracy derived from dominant features is slightly higher because the feature selection algorithm excludes noisy features leading to a better result. We report dominant features for each user in the last column of Table \ref{tbl:kmeans_accuracy}. We observed that the ringer mode is a dominant feature.% in all the users' data. 

In summary, this study shows that the designed \spyware can estimate and learn with an average accuracy of $90.3\%$  the user behavior, in particular: 
\begin{enumerate}[noitemsep, itemindent=!, nolistsep]
\item Average commuting time between home and work.
\item Average time spent at work and at home.
\item Weekend behavior, such as if a specific place is frequently visited on Sundays and average time spent at home.
\end{enumerate}

\subsubsection{\textbf{Experiment 2: Detection Using Current Antivirus Apps}}\vspace{-2mm}
 %To ensure the stealthiness, this spyware cannot catch attentions of any malware detection tool. 
After we had implemented this spyware app, we examined it using 5 well-known anti-virus applications \footnote{These 5 anti-virus applications are AVG AntiVirus, Symantec Norton Security \& AntiVirus, AVAST Mobile Security \& Antivirus, McAfee Security \& Power Booster, and Kaspersky Internet Security for Android.}. None of them reported this app as malware. This follows from the fact that the proposed \spyware does not have any suspicious code signature. This motivates the need to find a new detection technique that suits this kind of spyware.
%
%
%\begin{table}[t]
%\small
%\centering
%\begin{tabular}{|L{5.25cm}|C{2.5cm}|}
%\hline
%\textbf{Anti-virus Package Name} & \textbf{Scanning Result} \\ 
%\hline
%AVG AntiVirus 						& no threat \\ \hline 
%Symantec Norton Security \& AntiVirus	& no threat \\ \hline 
%AVAST Mobile Security \& Antivirus 		& no threat \\ \hline 
%McAfee Security \& Power Booster		& no threat \\ \hline 
%Kaspersky Internet Security for Android	& no threat \\ \hline \hline
%\end{tabular}
%\caption{Results of scanning the developed \spyware using signature-based malware detection packages. \vspace{-0.4cm}}
% \label{tab:antivirus}
%\end{table}
%
%
%
%\subsection{Experiment 3: Significance of \spyware}\vspace{-1mm}

\vspace{-0.1cm}
\subsubsection{\textbf{Experiment 3: Beyond Location \spyware}}
\label{sec:exp3}
%\r{Here choose some  apps that reads the settings (which we wrote in the discussion) and see the estimate MI from data we already collected (for example, app reads only ringer mode and ringer volume, calculate the corresponding estimate of MI from the data we have and see the corresponding accuracy from the data we have) then say that these apps can be an undetected malware}

%The \spyware we showed in this section uses the adaptation in the phone settings to leak user's behavior. 

 %In order to know the actual usage of these APIs in applications presented in the market and hence
%To understand the significance of \spyware, 

%\subsubsection{Beyond Location \spyware}

While the previous experiment aims at studying how the proposed \spyware can leak the semantics of the user location, we further explore how acquiring side-channels can reveal other sensitive user information. To this end, we study several context-aware apps in the Android market and report the monitored \emph{context} and the corresponding \emph{actions} taken by these apps in Table~\ref{tbl:contextappsidechannel}. Since other apps can observe these actions (even without asking for user permissions), these actions open a side-channel that leaks information about the user behavior. 
%For example, if a \spyware knows a priori the presence of Silence App~\cite{silenceapp} (an app which  changes your phone settings based on the calendar events), it can reveal the timing or repetition of calendar events based on the side-channel of phone settings. 
%For example, we can modify the \spyware developed in the previous user study 
%(which is used to leak location information through adaptations in phone settings)
%to leak other information like timing and repetition of calendar events whenever Silence~\cite{silenceapp} is used to adapt the phone settings according to calendar events.
For example, monitoring changes in the played music media\footnote{While Android framework does not provide an API to directly retrieve which music is playing, our experiments show that a \spyware can retrieve such information by reading the metadata associated with the currently active media.} can leak information about the user biometrics (heart rate and running pace) and user mood whenever such context-aware apps are used. In general, the proposed \spyware can take advantage of any pair of \texttt{get} and \texttt{set} methods that are present in the Android framework APIs.

\begin{table}[!t]
\small
\centering
\begin{tabular}{| C{3.5cm} || c | c | c |}
\hline 
Context-aware App                  & Context 	 &  Side-channel \\ \hline \hline
Tasker~\cite{tasker}  			& location   	 &  phone settings \\ \hline
Locale~\cite{locale}  			& location     	 &  phone settings  \\ \hline
%Silence~\cite{silenceapp}           & calendar events     &  phone settings  \\ \hline
%Alarm clock sleep cycle~\cite{sleepcycleapp}     & sleep cycle     & alarm clock    \\ \hline
RockMyRun~\cite{rockmyrunapp} & biometrics & music played \\ \hline
HABU music~\cite{habuapp}               & mood        &  music played \\ \hline
\hline
\end{tabular}
\caption{Context-aware apps in the market and their side-channel\vspace{-6mm}}
\label{tbl:contextappsidechannel}
\end{table}

\begin{figure*}[!t]
\centering
  \begin{subfigure}[b]{.33\textwidth}
    \centering
    \includegraphics[width=1\textwidth, trim={0 0 0 0},clip]{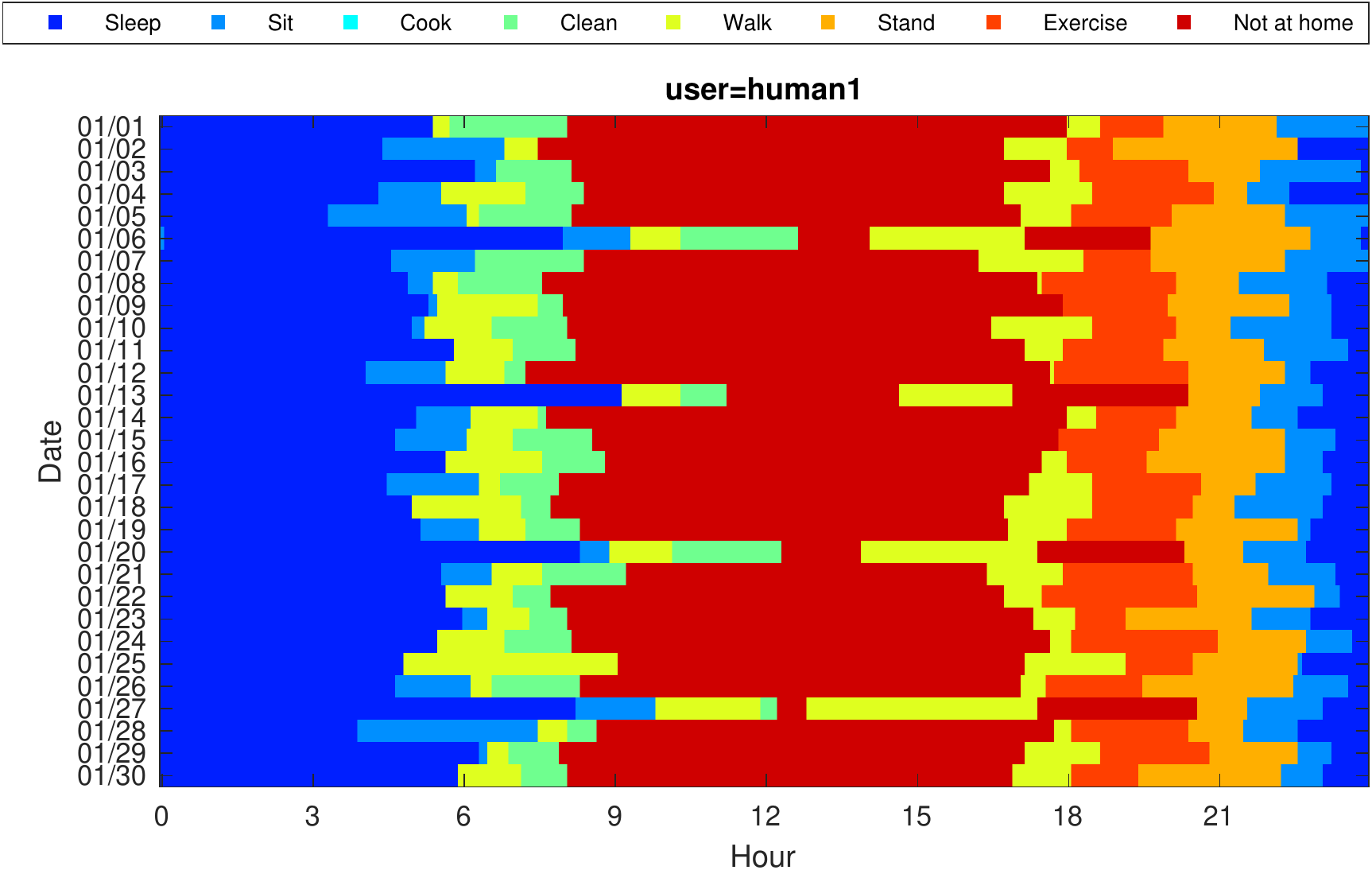}
    \caption{\small Golden output for human \#1 activity \label{fig:activity_timeline_golden}}
  \end{subfigure}
  \begin{subfigure}[b]{.33\textwidth}
    \centering
    \includegraphics[width=1\textwidth, trim={0 0 0 0},clip]{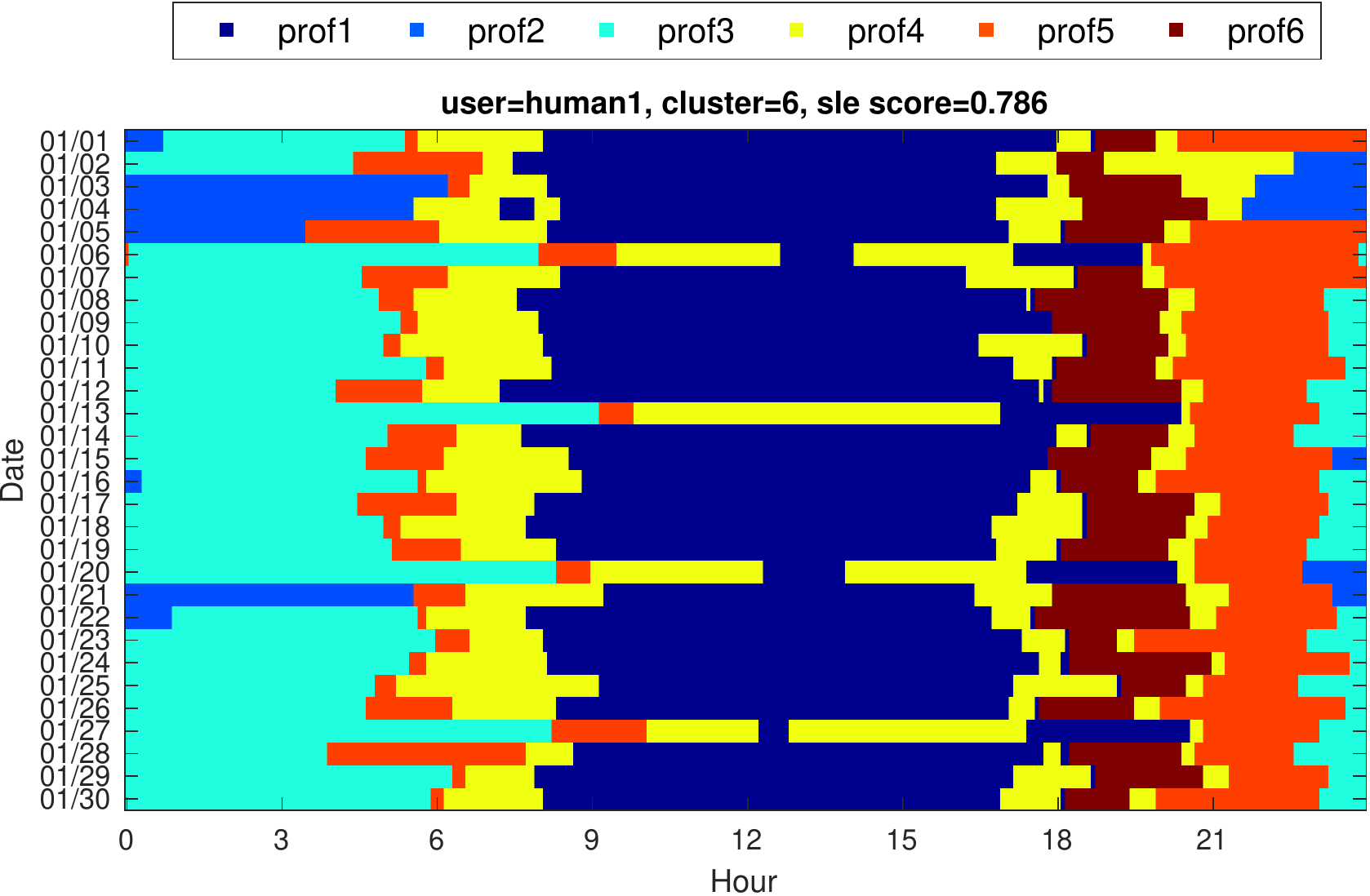}
    \caption{\small Clustered into 6 profiles \label{fig:activity_timeline_full}}
  \end{subfigure}
   \begin{subfigure}[b]{.33\textwidth}
    \centering
    \includegraphics[width=1\textwidth, trim={0 0 0 0},clip]{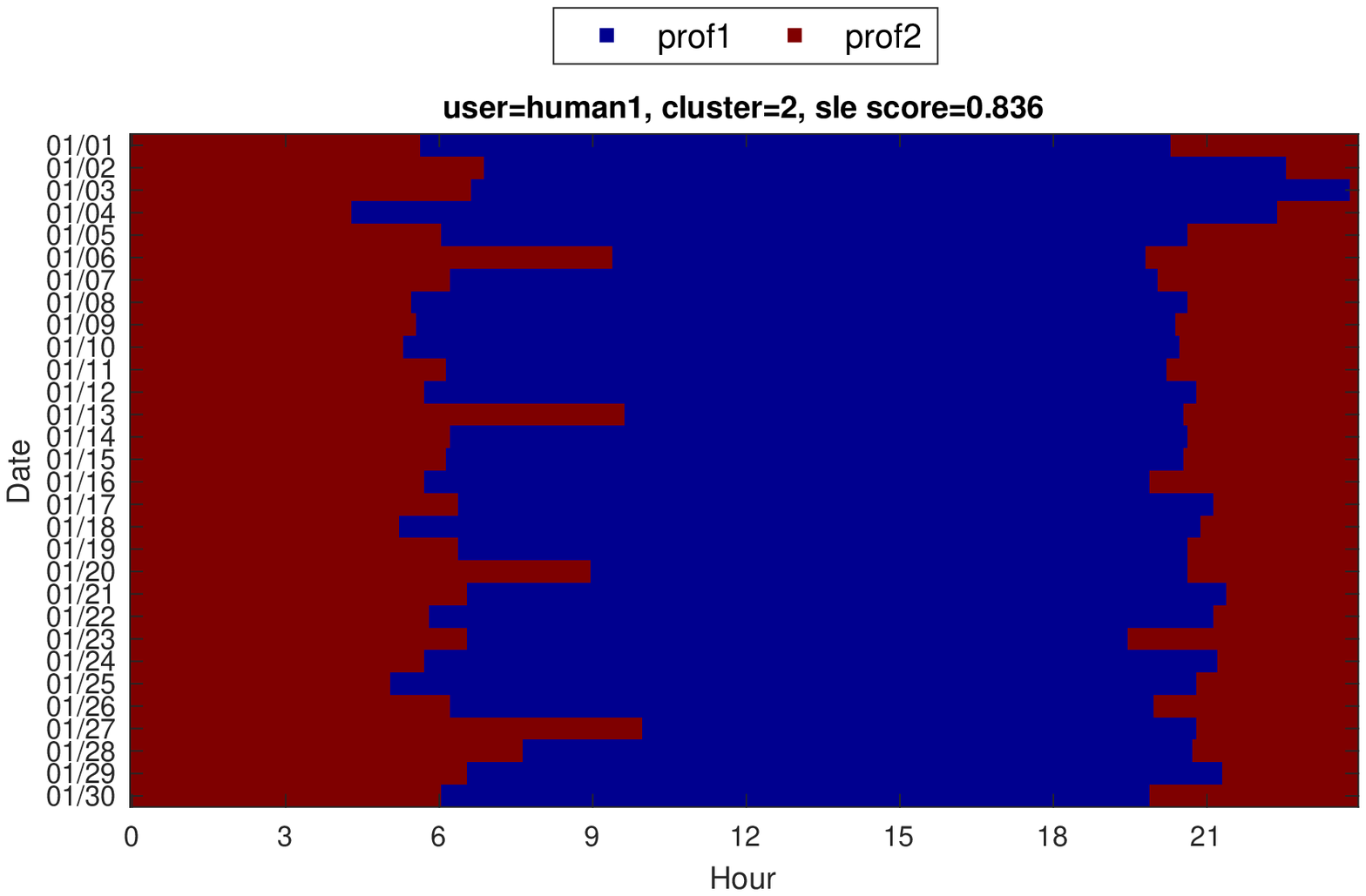}
    \caption{\small Clustered into 2 profiles  \label{fig:occupancy_timeline_full}}
  \end{subfigure}
\caption{Activity timeline from simulated human \#1. \label{fig:human1_timeline_example}\vspace{-0.1cm}}
\end{figure*}

\vspace{-2mm}
\subsection{\spyware Cloud: Smart HVAC System}
Cloud servers continue to be one of the weakest points when it comes to data breaches~\cite{almorsy2016analysis}. We argue in this example, that even non-sensitive information collected on HITL IoT clouds can be used to leak sensitive user information. To that end, we choose the personalized smart HVAC system as an example of a HITL IoT application~\cite{jung2017towards, hang2013carrying}. The personalized HVAC incorporate the human's activity to change the HVAC set point to maximize his comfort level~\cite{elmalaki2018internet}. In this scenario, as pictorially shown in Figure~\ref{fig:SpyConCloud}, human's activity such as cooking, sleeping, sitting, etc. is used with the current temperature in the house to tune the HVAC set point. The activity of the human is a wealth of information that should be kept safe as it leaks the behavioral pattern of the human along with his house occupancy patterns. Although the HVAC set point is calculated on the phone and pushed to the cloud service that controls the smart thermostat, a \spyware operating on
 the cloud could monitor the changes in temperature set points along with the current home temperature to infer the human's daily behavior.

%
 %However, as shown in Figure~\ref{fig:SpyConCloud}, a \spyware can be located at 

\vspace{-2mm}
\subsubsection{\textbf{Spyware Description}}\label{sec:spywarehvac}
We conduct a simulation-based experiment using EnergyPlus~\cite{crawley2000energy}, an industrial-level physics-based simulation engine, to model HVAC systems.
We use the weather reports in Colorado-Denver during January 2018~\cite{weatherforecaset}.  
The user activity is used to control the set point of the HVAC~\cite{elmalaki2018internet} across the day to maximize the human thermal comfort measured in Prediction Mean Vote (PMV)~\cite{fanger1970thermal}.
A \spyware mounted in the cloud can monitor the following information to infer the user's daily behavior: (1) Changes in the HVAC set point triggered by the IoT, (2) current house temperature, (3) time of the day, and (4) day of the week.

\subsubsection{\textbf{Experiment 1b: Data Mining by Clustering}}
 Using the same procedure in Experiment 1a (Section~\ref{sec:sca}), we simulated several humans independently in EnergyPlus. Due to space, we only show the ground-truth activity and the occupancy of one human (human \#1) across time for a month in Figure~\ref{fig:activity_timeline_golden}. The results of the clustering algorithm used by the \spyware to infer the human's daily behavior and the home occupancy are shown in Figure~\ref{fig:activity_timeline_full} and~\ref{fig:occupancy_timeline_full},  respectively. 

\paragraph{\textbf{Privacy Implications}} 
The results shown in  Figure~\ref{fig:activity_timeline_full} suggest that \spyware operating in the cloud can infer sensitive information like when the IoT user wakes up and goes to sleep (prof3), and when the user leaves the house and comes back (prof1). \spyware Cloud is also able to detect the occurrence of a periodic user activity just after returning home from work which is not performed during the weekends (prof6). The accuracy of the clustering is listed in Table~\ref{tbl:kmeans_accuracy_hvac}. By using two clusters to detect the occupancy (home/away), \spyware Cloud achieves an accuracy of $75\%$ for human \#1 and $91.72\%$ for human \#2. To detect the daily behavioral patterns, we increase the amount of clusters and achieved an accuracy of $76.2\%$ for human \#1 and $65.4\%$ for human \#2.

%Analyzing the results shown in  Figure~\ref{fig:activity_timeline_full}, we conclude that \spyware operating at the cloud can infer sensitive information like the times in which the IoT user wakes up and goes to sleep (prof3), along with the times he leaves the house and comes back (prof1). \spyware Cloud is also able to detect the occurrence of a periodic user activity just after returning home from work which is not performed during the weekends (prof6). The accuracy of the clustering (as defined in Section~\ref{sec:privacyImplication}) is listed in Table~\ref{tbl:kmeans_accuracy_hvac}. By using two clusters to detect the occupancy (home/away), \spyware Cloud achieves an accuracy of $75\%$ for human \#1 and $91.72\%$ for human \#2. To detect the daily behavioral pattern we use a different number of clusters. In this case, \spyware Cloud achieved an average accuracy of $76.2\%$ for human \#1 and $65.4\%$ for human \#2. 
%To emphasis, as mentioned in Section~\ref{sec:sca}, \spyware Cloud does not have access to ground-truth output to know its accuracy of detection. We only keep the golden output to show the privacy implication of the proposed \spyware.

 \begin{table}[!h]
\small
\centering
\begin{tabular}{| c || l | l | l | l | l | l |}
\hline 
 UID& \multicolumn{6}{c|}{Number of Clusters (k-means) }   \\
% & \multicolumn{4}{c||}{ } & \multicolumn{2}{c|}{features} \\
   \cline{2-7}
  & 2 (Occupancy) & 4 & 5 & 6 & 7 & 8   \\ \hline \hline
1 & 75.65 & 73.47 & 72.59 & 76.71 & 85.31 & 73.06\\ \hline
2 & 91.72   & 70.08 & 70.14   & 73 & 63.23 & 50.52\\ \hline
\hline
\end{tabular}
\caption{Clustering accuracy (in percentage) to detect the human's daily behavioral pattern for two simulated humans in a house using EnergyPlus simulator to simulate HVAC system. The clustering algorithm (k-means) uses features mentioned in Section~\ref{sec:spywarehvac}\vspace{-2mm}}
\label{tbl:kmeans_accuracy_hvac}
\end{table}

%\vspace{-0.2cm}
\section{\sysname System Architecture}\label{sec:sysarch}
In \sysname, we focus on the general question of how to design software mechanisms that can detect and mitigate \spyware. Our entry point is that mobile phones play an important role in connecting the user to numerous HITL IoT. For example, in both the scenarios discussed in Section~\ref{sec:spycon}, namely when a \spyware operated on the edge/phone and the cloud, sensitive user information are collected from the user's phone. Being the source of several user's sensitive information, a natural choice of implementing a privacy safeguard would be on the user's phone. Indeed, the same concepts developed in this section can be applied to other sources of sensitive user information before it is used by the rest of the HITL IoT.
%While the ``Phone Manager \spyware'' spyware described in Section~\ref{sec:spycon}  is just one example of how the very act of adapting to user context leaks sensitive information, in the remainder of this paper, we focus on the general question of how to design software mechanisms that can detect and mitigate such spyware. 
%\begin{figure}[t]
%  \centering
%    \includegraphics[width=\columnwidth, trim={1cm 9.6cm 0.5cm 1.1cm},clip]{figures/generic_architecture_v2}
%	\caption{A conceptual architecture of \sysname showing the three main components: context-adaptation registration module, information-based detection engine, and mitigation engine.\vspace{-5mm}}  \label{fig:archdetect}
%\end{figure}

%\vspace{-0.2cm}
\subsection{Threat Model and Design Objectives}
Motivated by the observations from the previous section, we define our threat model as follows.\\
\noindent\textbf{[T1] No prior signature or behavior}: Since \spyware is a new class of spyware, we assume neither prior knowledge of code signatures nor prior recorded suspicious behaviors exist.\\
\noindent\textbf{[T2] Unknown inference algorithm}: We assume no knowledge of the algorithms used by \spyware to infer the sensitive information. \\
\noindent\textbf{[T3] Access to information}: We assume that \spyware has access to the APIs\footnote{Our study in the section~\ref{sec:spyconedge} shows that many of such Android APIs do not require permissions.} on the phone   and have access to the information stored in the cloud.\\ %Our study in the previous section shows that many of such Android APIs do not require permissions.\\
\noindent\textbf{[T4] Knowledge of Existence of \sysname}: Finally, we assume that \spyware is aware of the existence of \sysname. That is,  \spyware has full knowledge of \sysname detection and mitigation algorithms and \spyware may adapt its behavior accordingly.
%we assume that \spyware knows the exact algorithms used by \sysname and adapts its behavior accordingly.

T1 and T2 imply that traditional malware detection schemes (e.g.,~\cite{xie2010pbmds}) are not adequate in our problem setup.
%Traditional detection schemes of malware are based on detecting certain signatures in the malware code. While the signature-based malware detection is used extensively on non-mobile computing platforms, it was recently shown that signature-based detection is neither real-time nor independent on users' awareness~\cite{xie2010pbmds}. Instead, behavioral-based malware detection schemes~\cite{xie2010pbmds} have the advantage that they ``learn'' new malware behaviors and mitigate them accordingly.  Unfortunately, neither the signature-based nor the behavioral-based schemes fit the problem under consideration in this work. This follows from the fact that both signature-based and behavioral-based schemes need prior information of previous malware in the form of either malware signature or malware behavior. Since neither a signature nor the behavior of \spyware does exist so far, such schemes can not be used. 
Hence, a new \spyware detection scheme has to be designed without possessing any prior information (signature or behavior) of such spyware. In particular, we propose the following design objectives in \sysname:\\%\vspace{2mm}% ~\\
\noindent\textbf{[O1] Generic Detection}: \sysname should be able to detect any possible privacy leak without prior knowledge of spyware signatures or behavior. \\%\vspace{2mm}% ~\\
%-based applications should add minimal overhead on the application developer at developing time;
\noindent\textbf{[O2] Mitigation}: \sysname should be able to mitigate the impact of a possible \spyware while minimally affecting the performance of the HITL IoT.  \\%\vspace{2mm}% ~\\
\noindent\textbf{[O3] Performance}: \sysname should be lightweight and add minimal execution overhead.

%\r{The objectives need to be rewritten}

Motivated by these three design objectives, we designed \sysname as discussed in this section in three main blocks: Context-adaptation registration, Detection Engine, and Mitigation Engine.  % A conceptual overview of \sysname architecture can be shown in Figure~\ref{fig:archdetect}. 
%In the remainder of this section, we discuss the overall architecture of \sysname. 
The details of each building block are given in the subsequent sections.
\vspace{-4mm}
\subsection{Context-adaptation Registration}\label{sec:contadaptreg}
In order to protect the actions and decisions 
%adaptions 
made by the HITL IoT, 
%context-aware apps, 
\sysname needs to know the context-action relations within the IoT system. %apps. 
%systems monitor context and take adaptation actions accordingly. %based on the reported values from the current context. 
%\sysname needs to learn these context-action relations. % in order to shield the information leakage from a possible \spyware.
One action in an adaption can be made based on several context, and one context can be shared by different actions.
For example, a user may use a camera app and another app for listening to music in the background while running. Upon detecting that the user is running, the camera app adjusts the hardware camera focus while the music app changes the music playlist. Both actions from these two apps, i.e., changing camera properties and altering music playlist, are made based on the same context, i.e., user's physical activity. Providing the context-action relations is essential for \sysname to monitor how likely a \spyware is inferring associated context. %which depends on the registration files.

%The scope of the paper is \emph{not} to understand the semantics of the source code of a context-aware application but rather to emphasize the possibility of side channel attack by \spyware and to demonstrate the protection against \spyware. With that position in mind, we choose a simple design model to facilitate the extraction of context-action relations. 
 In order to automatically retrieve  the context-action relations, we use FlowDroid~\cite{arzt2014flowdroid} (which augments the Java code analysis tool Soot~\cite{sootframework}) in order to analyze the call flow graph of the context-aware app (e.g., Tasker) and extract all relations between the setter APIs and getter APIs along with the information about which data will be sent and recieved over the network interface to the cloud. The final output is an XML file that maps context (getter APIs and data received from the cloud) to actions (setter APIs and data sent to the cloud). We call this XML file a \emph{registry file}. The developer can then view the automatically generated XML file to verify or modify it before being used by \sysname.
 
In the registry file example shown in Figure~\ref{fig:registry_example_2}, there are two \texttt{adaptation} policies. Each \emph{adaptation} represents one context-action relation. The first adaptation %is relatively straightforward, which 
makes decisions based on GPS location and updates ringer mode and alarm volume accordingly. In the second adaptation, the application adjusts the camera settings to accommodate battery capacity, GPS location, and transportation modality.

%Since the developers of the context-aware application understand the semantics of the app, we ask developers to provide a \emph{registry file} formatted in XML that maps context to actions within the context-aware app.
%However, understanding the semantics of code to extract the context-action relations can be achieved using source code analyzers like lint~\cite{androidlint} or by analyzing the call flow graph of the decompiled Android application using tools like SAAF~\cite{hoffmann2013slicing}. 

During application installation phase, \sysname checks the existence of the registry file. %inside the application package 
%APK.
If the file exists, \sysname considers the associated application as a context-aware app and will offer detection and mitigation mechanisms to protect this app from potential context exposures against any other suspicious \spyware. %We describe the structure of the registry file as well as the post-processing steps performed by \sysname below. %over this file.

%\vspace{-2mm}
%\subsubsection{\textbf{Registry file structure}}\vspace{-1mm}
%A context-aware application typically invokes a set of getter APIs for acquiring the latest \emph{context} and a set of setter APIs to perform the corresponding \emph{actions}. %In an application, different actions can be adapted based on different sets of contexts. %Therefore, developers can specify several \texttt{adaptation} elements in a registry file to define context and action correspondences. %Thus, each adaptation should contain exactly one \texttt{context} and one \texttt{action} tag, both include one or more methods which are going to be used in the context-aware app code.
%%Figure~\ref{fig:registry_example_2} demonstrates an example of the registry file.
%In the registry file example shown in Figure~\ref{fig:registry_example_2}, there are two \texttt{adaptation} policies. The first adaptation %is relatively straightforward, which 
%makes decisions based on GPS location and updates ringer mode and alarm volume accordingly. In the second adaptation, the application adjusts the camera settings to accommodate battery capacity, GPS location, and transportation modality. %Note that both actions have to be embedded in two separate adaption policies although these two actions are conducted (partially) based on GPS location context because the decisions (i.e., actions) are computed independently. 

\begin{figure}[!t]
\centering
  \lstset{basicstyle=\scriptsize, language=XML, morekeywords={adaptation,action,context}}
{
\begin{lstlisting}
<adaptation>
	<context> getGPSlocation() </context>
	<action>  setRingerMode()  </action>
	<action>  setAlarmVolume() </action>	
</adaptation>
<adaptation>
	<context> getTransportationModality() </context>
	<context> getBatteryCapacity()        </context>
	<context> getGPSlocation()            </context>
	<action>  setFocusMode()              </action>
	<action>  setJpegQuality()            </action>
	<action>  setSceneMode()              </action>
</adaptation>
\end{lstlisting}}
  \caption{Snippet of a \sysname registry file. \vspace{-0.5cm}}\label{fig:registry_example_2}
\end{figure}

\vspace{-2mm}
\subsection{Information-Based Detection Engine}
\label{sec:detection}
%As discussed earlier in this section, neither signature-based nor behavioral-based detection is suitable for detecting \spyware malware. Instead, we propose an information theoretic detection scheme. In this scheme, we measure the \emph{amount of information} possessed by any suspicious application. By relying on information-theoretic measures, our detection engine is agnostic to the learning algorithm that is used by the malware. We refer to this scheme as \emph{information-based} detection scheme. 

%Malware detection techniques are traditionally based on a \emph{prior knowledge} of either the code signature or the behavior of \emph{similar} applications that are previously marked as spyware. However, 

%As we showed in Section~\ref{sec:spycon_app}, 
It follows from assumptions T1 and T2, in our threat model,  that existing signature-based detection mechanisms are not able to detect the developed \spyware. Similarly, behavior-based detection techniques are not suitable to detect such spyware since no prior behavior is known. Exacerbating the situation,  the behavior of \spyware is coupled to the behavior of the authentic context-aware apps that are part of the HITL IoT. That is, if the IoT (and hence the authentic context-aware app) triggers actions more frequently, \spyware will monitor the actions by calling the getter values APIs for these actions more frequently. This behavior coupling hinders the usability of the behavior-based detection techniques.

The idea behind the information-based detection is to keep track of the ability of \spyware to infer the context through monitoring actions triggered by this context. Recall that we do not have any prior knowledge or assumption on how \spyware performs its inference (T2). To this end, we draw on the literature of information theory and leverage \emph{mutual information} to quantify the amount of correlation (or dependence) between two random variables. In our scenario, we use the mutual information between \emph{context} and \emph{action} as a metric to measure how certain a \spyware can infer context from observed actions. Mutual information provides a theoretical bound on the inference capability of \emph{any} learning algorithm. 
Generally speaking, the lower the mutual information between context and actions is, the lower the accuracy \emph{any} inference algorithm can get.
Push into one extreme, if the mutual information is zero, then \emph{no} algorithm can infer context from monitored actions.

Based on this intuition, our detection engine constantly tracks what actions have been monitored by each app, computes an estimate of mutual information between the actual context and those actions monitored by this app, and assigns a \emph{suspicion score} accordingly. This score is then passed to our Mitigation Engine, as an indicator of the magnitude of countermeasures, which aims at reducing the amount of information possessed by suspicious apps. 

\vspace{-0.2cm}
\subsection{Mitigation Engine}\label{sec:mitigation}
Once the \emph{suspicion score}, assigned to each possibly running \spyware, increases beyond a certain threshold (named alarm threshold), the Mitigation Engine informs the user with the possibility of having a \spyware and
asks the user permission to apply countermeasures against the suspicious app. Upon the user consent, the Mitigation Engine seeks a general way to hinder the \spyware's capability of revealing user context. While completely blocking all side-channels may not be practical, our goal is to drastically reduce its bandwidth.  This process needs to be done without any prior assumption on the type of inference algorithms used by \spyware (T2). There are two solution regimes to achieve this: 1) imposing delays so that a \spyware cannot get the latest action updates in real time, and 2) tearing down the correlation between the actions monitored by \spyware and the associated context. We call the mitigation method in the first regime \emph{delay} and introduce three different mitigation methods in the second regime: \emph{suppression}, \emph{row-masking}, and \emph{feature-masking}.

Before delving into the details of various mitigation algorithms, we would like to stress the following facts:
\begin{itemize}
\item \textbf{Complementary and adaptive mitigation:} The two regimes of mitigation methods achieve different goals and can complement each other. In practical scenarios, we combine both delay mitigation with one in the second regime which reduces the mutual information. \sysname does not assume any mitigation can always outperform the others. In fact, a mitigation technique may be effective only in a certain situation. Hence, \sysname starts by selecting a mitigation method and applies it accordingly. Next, \sysname
 tracks the effectiveness of the imposed mitigation technique by constant monitoring of changes in the 
 suspicion score.
If a \spyware can still survive under current mitigation treatment, 
%\r{i.e., the suspicion score still exceeds a predefined threshold (SE: threshold was never mentioned before or after),} 
\sysname increases the mitigation magnitude first and eventually switches to a different method.
\item \textbf{The Scope of mitigation:} The mitigation is applied on a per-app basis instead of a system-wide configuration. Hence, well-behaved apps (including the protected context-aware app) can receive correct action values. Therefore, mitigations cause no negative impact on context-aware apps.
%\r{do we need to emphasize the difference between mitigation against Edge Spycon vs Cloud SpyCon?}
\end{itemize}

The rest of this section we explain the details of the mitigation techniques we have in \sysname.
%The proposed mitigation techniques can be defined as follows:
\subsubsection{\textbf{Mitigate by Delay}}\label{sec:delay}
\vspace{-2.2mm}
%Assume that a \spyware keeps polling changes in actions in order to detect when a certain adaptation takes place.
%As shown in Section~\ref{sec:sca}, monitoring the time of the day with the changes done by context-aware app can reveal user's schedule.
%As shown in Section~\ref{sec:sca}, monitoring the phone setting changes updated by the context-aware app can reveal the semantics of the user location and thus infer user's schedule.
Our first strategy is to let \sysname \emph{delay} the information about when adaptation actions are taken so that a \spyware can only get an outdated context instead of the latest one. That is, whenever an app with high suspicious score or whenever the cloud is getting information about actions taken based on changes in user contexts, \sysname will send old information that are delayed 
%\sysname chooses a delay of 
$d$ minutes which is selected randomly (to prevent the \spyware from revealing the delay). % \sysname keeps a history of $d$ minutes of all monitored actions so that getters are kept to be unblocking calls.
%As a consequence, a \spyware can still reconstruct user context sequences but with unknown offset. \sysname increases the delay adaptively based on how suspicious a \spyware is.

%\begin{figure}[t]
%  \centering
%    \includegraphics[width=\columnwidth, trim={0cm 12.5cm 0.5cm 0cm},clip]{figures/architecture5}
%     \caption{\sysname architecture. The context-aware application is registered in \sysname by context-adaptation registration module. The behavior of the context-aware app is monitored, and a possible \spyware is detected by the Detection Engine. Adequate mitigation technique is then taken by the Mitigation Engine\vspace{-5mm}%\sysname service runs in the system space while its app interface (\sysname manager) runs in the user space. %Context-aware application is registered in \sysname by context-adaptation registration module. The behavior of the context-aware app is monitored and a possible \spyware is detected by the Detection Engine, then the adequate mitigation technique is taken by the Mitigation Engine.
%    }
%     \label{fig:androidarch}    
%\end{figure}

\begin{figure*}[!t]
\centering 
\includegraphics[width=0.95\textwidth, trim={0 16cm 0 0},clip]{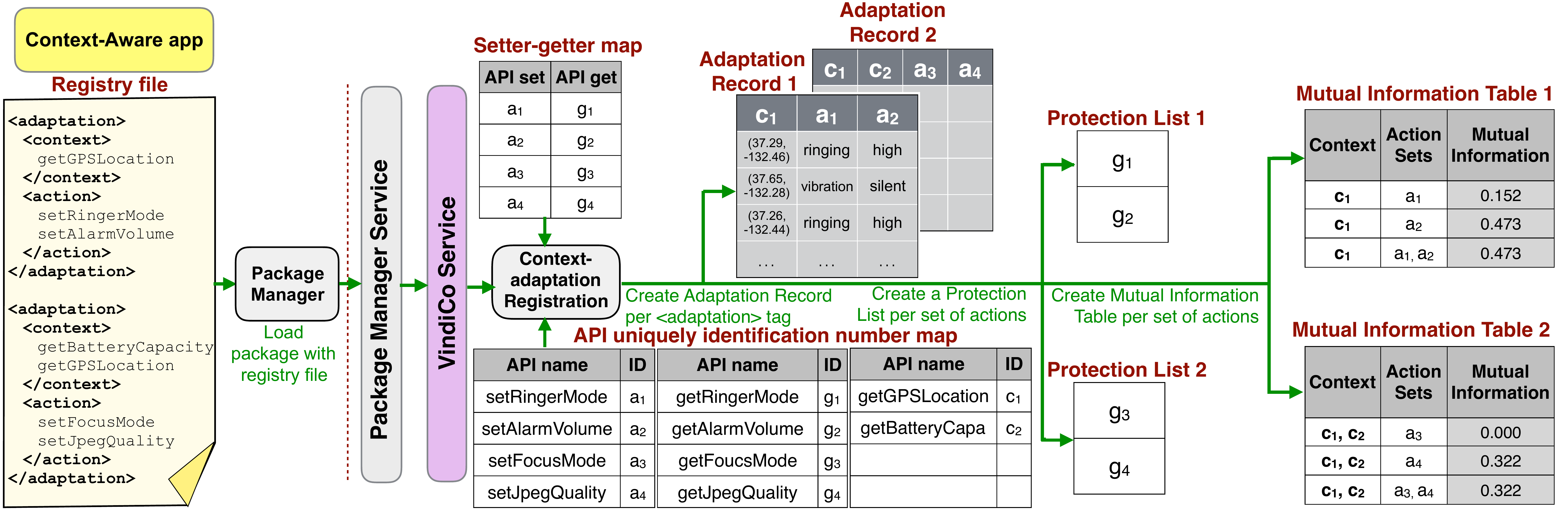}
  \caption{Architecture and data structures used by \sysname. At installation time, \sysname checks the existence of a registry file and starts processing it accordingly. Registry file processing constructs all necessary data structures that are needed  by the detection and mitigation engines \vspace{-5mm}}\label{fig:registryfileprocess}
\end{figure*}

\vspace{-2.2mm}
\subsubsection{\textbf{Mitigate by Suppression}}\label{sec:suppression}
With the previous mitigation method, even though \spyware may not be able to get the real-time update from delay mitigation, a \spyware can retrofit users' daily routine by aligning with a \emph{priori} knowledge, for example, when a person usually goes to school or work. Once the personal schedule is derived, this mitigation will have no effect in the future. The fundamental reason is that simply imposing a delay will not reduce the mutual information. Suppression mitigation, on the other hand, is designed to decrease the information a \spyware can harvest. The main idea of this mitigation technique is to give \spyware false adaptation actions. For \spyware that are detected on the phone (i.e., associated with high-suspicious score), \sysname can return action values associated with another context without affecting the actual actions taken by the trusted app and hence eliminating any negative affect on the system. On other hand, if the cloud is associated with a high-suspicious score, \sysname will corrupt the actions before they are sent to the cloud leading to some negative effect on the whole system performance. 
%When a \spyware is detected and suppression mitigation is chosen to be applied,
% \sysname suppression mitigation module randomly chooses one in the latest $k$ recorded values for the action. %which we call a record.
 % The number $k$ is the mitigation magnitude of this method.

\subsubsection{\textbf{Mitigate by Masking}}\label{sec:mitigation_mask}
Yet another mitigation to increase obfuscation is to mask some action values, i.e., returning zeros. We explore two variants of the masking approach:  \emph{row-masking} and \emph{feature-masking}. In row-masking, our system returns correct action values, but with a certain probability $p$, our system returns 0 to \spyware for all the action values after context changes. The masking effect takes place until the next adaptation is made. In consequence, \spyware cannot infer anything during the masked periods, but can still make inferences based on the unmasked observations.
The feature-masking approach, in contrast, considers each action value individually. Upon an adaptation is made, each action has a certain probability to be masked. As a result, at any given context change, the \spyware can only observe partial action values after an adaptation occurs. % thus user context is less likely to be revealed. 
The decision of masking all actions in row-masking or which action values to be masked in feature-masking depends on flipping a biased coin with a selected probability $p$, which serves as the parameter of mitigation effectiveness.
The evaluation of the mitigation techniques is shown in Section~\ref{sec:evalmitigation}.

\vspace{-0.2cm}

\section{Implementation}
\sysname is implemented by modifying the Android system image for platform 5.0.1 API 22~\cite{aosp}. We extend the Android system layer to add the three main parts of  \sysname as shown in explained in Section~\ref{sec:sysarch}.  \sysname increases the sizes of system image by 4.5\% (original image size is 998 MByte). The overhead on the API call is shown in Section~\ref{sec:analysis}

 Upon installation time (as shown in Figure~\ref{fig:registryfileprocess})  \sysname searches for the registry file  and creates the necessary data structures to track all changes in user contexts and actions associated with these changes. It then creates a list called \emph{Protection Lists} and \emph{Mutual Information tables} that keeps track of the available information at each app and the cloud. To update these tables, \sysname intercepts all the calls between apps and the Android OS. Whenever an app (or the cloud) gets information about context changes, \sysname  starts to suspect this app (or the cloud) by assigning a \emph{suspicion score} to this app (or the cloud service).
  The value of this \emph{suspicion score} is
  equal to the mutual information associated with the \emph{getter} API (invoked by the app) which is stored in the \emph{mutual information tables}.
Since the same app may have called several \emph{getter} APIs belonging to different \emph{Protection Lists}, \sysname associates a set of \emph{suspicion scores} (instead of just one \emph{suspicion score}), one per \emph{Protection List} in a table that is called \emph{applications suspicion scores} table. %(see Figure~\ref{fig:detectengine}).
\subsection{\textbf{\sysname Detection Engine}}
 An important aspect that is related to the use of mutual information  is the convergence of estimated mutual information to the actual mutual information from the data samples accessed by \sysname. That is, it is well known that calculating an estimate of mutual information with enough precision requires the access to multiple samples from both context and triggered actions\footnote{An estimate of the mutual information converges to actual mutual information asymptotically.}. Hence, one can argue that a malicious spyware---who is aware of the existence of \sysname as per assumption T4 in our threat model---may be tempted to reduce the frequency for which it monitors the actions with the goal of decreasing the number of samples used to estimate his mutual information and hence \emph{deceives} \sysname by reducing the associated \emph{suspicion score}\footnote{On average the actual mutual information is the upper bound for the estimate mutual information.}. 
To avoid such situation, we associate mutual information to actions instead of apps. That is, whenever an action takes place, we update the amount of mutual information between this particular action and all the context associated with it.  At any time point, once an app monitors this action, we copy the mutual information associated with this action to the app. Therefore, regardless of how frequent an app monitors actions, we associate to it the mutual information based on all samples in history---recall that the number of these samples are not controlled by the malicious app---not only on those to which it has access.%~\\
\vspace{-2mm}
\subsection{\textbf{\sysname Mitigation Engine}}
As mentioned in Section~\ref{sec:mitigation}, \sysname does not presume any mitigation treatment outperforms the others, and opportunistically selects a method and see the effect and gradually increases the mitigation. 
 If the suspicion score cannot be effectively decreased, \sysname will switch to another mitigation method.  To optimize the run-time of mitigation engine, we cache the decisions taken by the mitigation engine on this API call for each calling application while a separate thread update this cache based on mutual information values.

\subsection{\textbf{Timing Analysis of \sysname}}\label{sec:analysis}
%We evaluate the timing performance of \sysname on Nexus 5 phones by running a modified system image for platform 5.0.1 API 22~\cite{aosp}. 
The execution time is obtained using Android traceview~\cite{trace}. 
%In order to reduce the performance overhead of \sysname on each API call, the Detection Engine is running in parallel thread. On the other hand, the Mitigation Engine runs on the same thread of any possible \spyware app.
%Hence whenever a set API is called as shown in Figure~\ref{fig:detectengine}, the call for the \sysname registry file processing and detection modules run independently to the call of the API. 
Table~\ref{tbl:timinganalysis} shows the CPU execution time when the complexity of the context-aware app increases, which is majorly reflected by the number of adaptation tags in the registry file. 
Parsing and processing the registry file take approximately $6 ms$. Fortunately, this overhead takes place only during the installation of a new package on the phone. On the other hand, at runtime, \sysname adds negligible overhead (less than $0.1 ms$) which is approximately $3\%$ increase from the average execution time of the original API call.
\begin{table}[!h]
\small
\begin{tabular}{| C{2.7cm} | C{1.2cm} | C{1.2cm} | C{1.2cm} | %L{10cm}|
}
\hline 
  \textbf{Description} & \multicolumn{3}{c|}{\textbf{Overhead in CPU Time (ms)}} \\ \cline{2-4}   %&
   % & \multicolumn{3}{c|}{} \\ \cline{2-4}   
  & \textbf{2 adapt} & \textbf{3 adapt} & \textbf{4 adapt} 
  %&   
  \\ \hline \hline
 
Parsing registry file & $3.186$  &  $3.186$  & $3.187$ 
%& \scriptsize{Overhead takes place during the installation time of the application.} 
\\ \hline

Registry file processing &$1.48$ & $2.22$ & $2.96$ 
%& \scriptsize{Construction of the necessary data structures. % in the context-adaptation registration module. 
%Overhead takes place while loading the application for the first time after the application is launched.} 
\\ \hline

API call in context-aware apps & $0.066$  &  $0.076$ &  $0.076$ 
%& \scriptsize{Tracking the values of context and action from context-aware application and filling the \emph{Adaptation Records} at runtime. Overhead takes place whenever an action API is called  by the authentic context-aware application. On average, the original API call consumes $2.56$ ms, and hence the overhead is $2.9\%$.} 
\\ \hline

API call in other apps &$0.056 $ & $0.090$  & $0.095$ 
%& %Overhead due to tracking different API calls in the \emph{Protection List} and extracting the corresponding mutual information. 
%\scriptsize{Overhead when a \texttt{get} API is called from one of the \emph{Protection List} by other apps. On average, the original API call consumes $2.56$ ms, and hence the overhead is $3\%$.}  
\\ \hline

\end{tabular}
\caption{Timing analysis of \sysname against increasing complexity of context-aware apps measured by number of adaptation tags in the registry file.\vspace{-0.3cm}}
\label{tbl:timinganalysis}
\end{table}

 \begin{figure*}[!ht]
\centering
\includegraphics[width=0.96\textwidth]{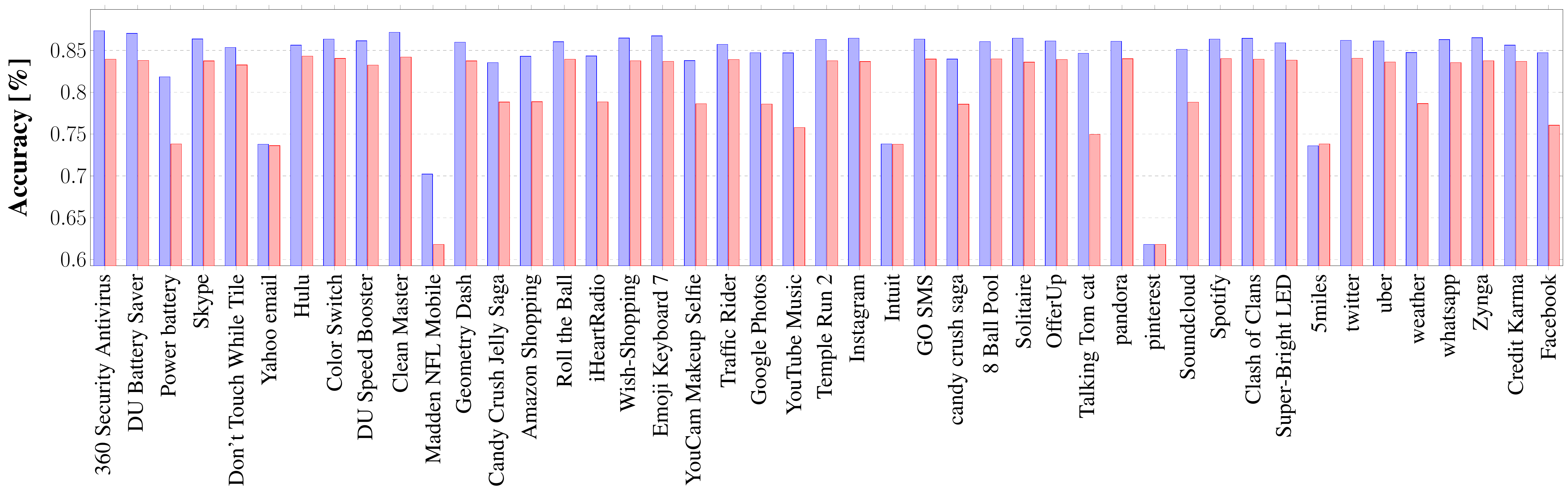}
\caption{Accuracy of leaking sensitive information about phone user from the data collected by
45 of the most downloaded free apps; (blue) accuracy of identifying the semantics of the user location when a location-based context-aware app (Tasker/Locale) is used to change the phone settings and (red)  accuracy of identifying user calendar profile when a calendar-based context-aware app (Silence 2.0) is used to change the phone settings. \vspace{-0.4cm}}
\label{fig:accuracy_apps}
\end{figure*}

\newpage
\section{Evaluation}\label{sec:evaluation}
In this section, we discuss 
%the implementation of \sysname along with showing
%the performance of our implemented \spyware. In particular, we study 
the performance of the proposed \sysname in mitigating possible SpyCon Edge and \spyware Cloud. %detection and mitigation methods by comparing the accuracy of the developed \spyware before and after applying the mitigation algorithms followed by a timing analysis of the implemented \sysname.

\vspace{-0.15cm}
\subsection{Experiment 4: How many SpyCon Edge in the market?}
We start by examining how many apps currently in the market and have the capability of being a successful \spyware Edge app. While identifying whether a real app is actually exploiting the proposed side-channel is hard to be checked without the knowledge of the app behavior, we focus in this experiment, instead, on identifying real apps in the market that possess enough information to leak sensitive information about phone user. 

To that end, we performed static analysis on 45 of the most downloaded free apps from the Google Play store to check which getter APIs are being used by these apps. Next, we intercepted these getter APIs %(using a similar technique to the one discussed in Section~\ref{sec:suspicious_score}) 
to retrieve the information possessed by these applications. Finally, we use this data to leak information about the phone user when a location-based context adaptation apps like Tasker/Locale and Silence (a calendar events-based context adaptation app) are taking actions by changing the phone setups based on location and calendar events, respectively. As mentioned in Section~\ref{sec:spycon}, context-based apps are increasingly used by IoT platforms like IBM Watson IoT to allow the designed IoT system to adapt its functionality to the human user.

Figure~\ref{fig:accuracy_apps} shows the accuracy of retrieving the user context from the data available to the 45 apps. Our results show that $86\%$ of the top downloaded free apps have enough information to leak user context with some of the apps scoring more than $80\%$ accuracy in leaking user context when a location-based context aware app is used. Similarly, for calendar-based context, $64\%$ of these apps can leak sensitive information with an accuracy more than $80\%$ showing the significance of this side-channel information.\vspace{-2mm}
\begin{figure}[!ht]
\centering
\includegraphics[trim={0 5cm 0cm 5cm}, clip, width=0.8\columnwidth]{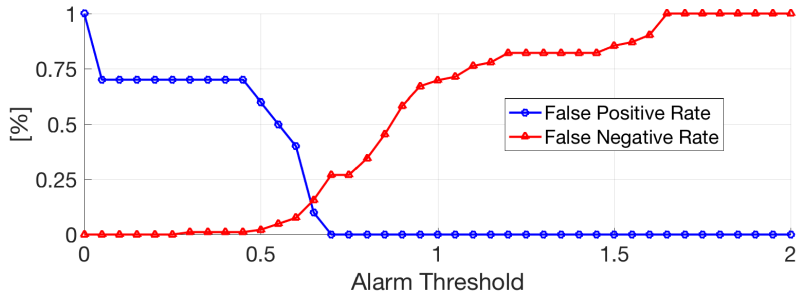}
\caption{Performance of the \sysname information-based detector (in terms of false positive and false negative rates) versus different alarm thresholds. \vspace{-0.5cm}}
\label{fig:fpfn}
\end{figure}

\vspace{-0.15cm}
\subsection{Experiment 5: Performance of Information-Based Detection}
\subsubsection{\textbf{Detection Accuracy}}
We investigate the effect of choosing the \emph{alarm threshold} (the threshold on the \emph{suspicion score} above which an app is considered malicious) on the performance of the proposed information-based detection algorithm. In particular, we evaluate the performance of the detection engine regarding both the false positive rate and false negative rate. In this context, an application is considered \emph{malicious} whenever its clustering accuracy exceeds the \emph{baseline accuracy} by blind guesses defined in Section~\ref{sec:privacyImplication}. A false positive flags the case when the information-based detector claims that an application possesses enough information to accurately identify the user behavior with accuracy more than the \emph{baseline accuracy} while the real accuracy of the app is indeed lower than the \emph{baseline accuracy}. A false negative is defined similarly. 
Similar to the previous experiment, we focus again on the case when \spyware is monitoring phone settings changed by a location-based context-aware app (Tasker/Locale) and when these settings are changed by a calendar events-based context-aware app (Silence). 

\begin{figure*}[!h]
\centering
  \begin{subfigure}[b]{.32\textwidth}
    \centering
    \includegraphics[width=1\textwidth]{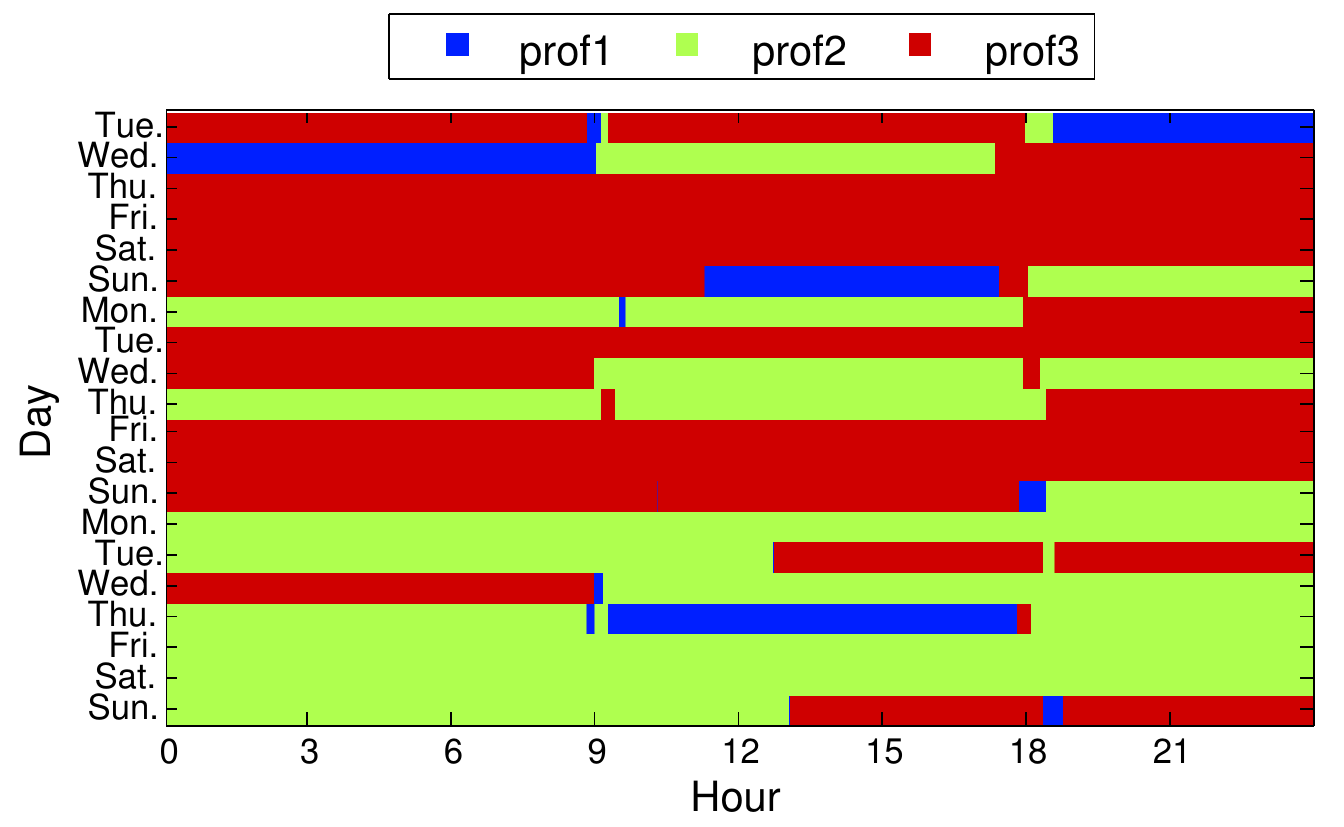}
    \caption{\small Suppression mitigation \emph{(3 rows)}}
    \label{fig:profile_timeline_miti_suppress}
  \end{subfigure}
  \begin{subfigure}[b]{.32\textwidth}
    \centering
    \includegraphics[width=1\textwidth]{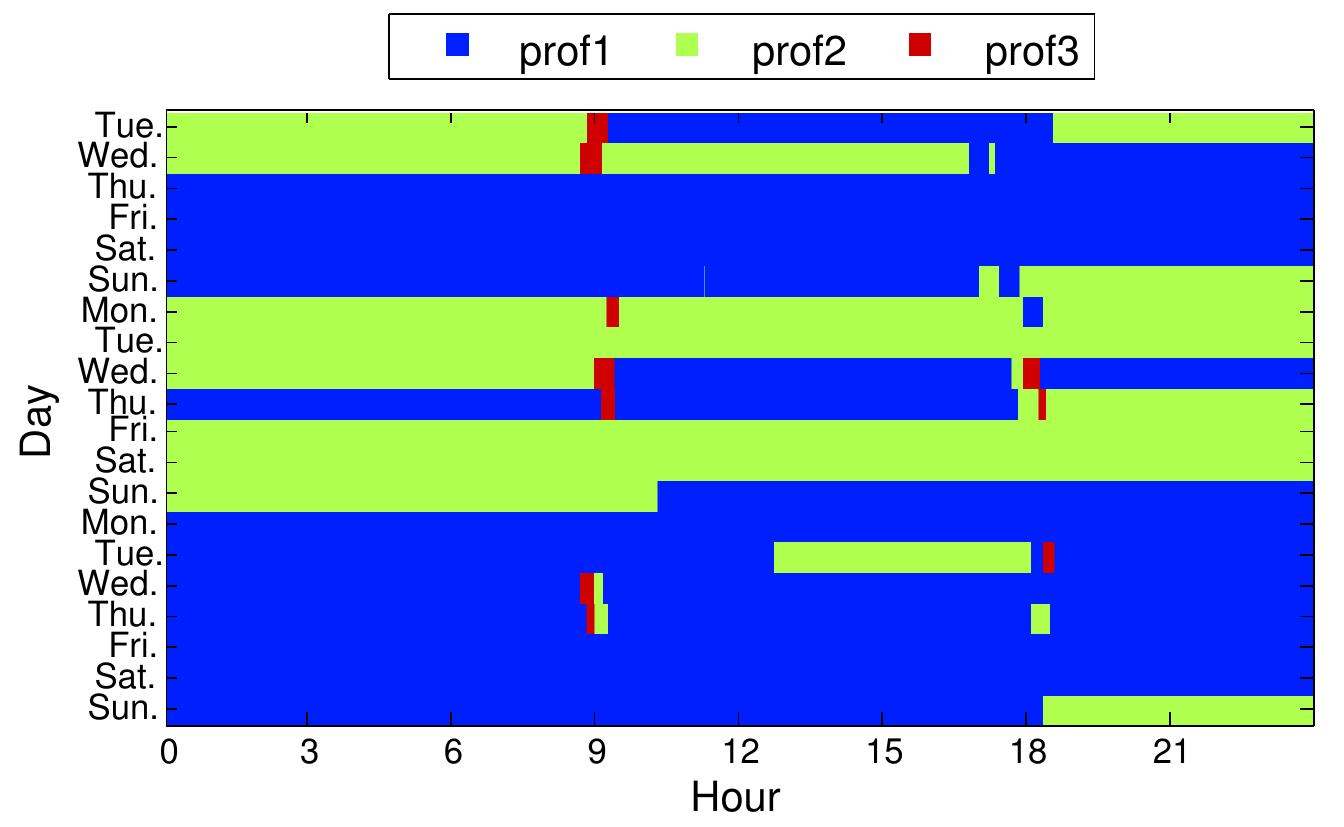}
    \caption{\small Row-masking mitigation \emph{(p=0.4)}}
    \label{fig:profile_timeline_miti_maskrow}
  \end{subfigure}
  \begin{subfigure}[b]{.32\textwidth}
    \centering
    \includegraphics[width=1\textwidth]{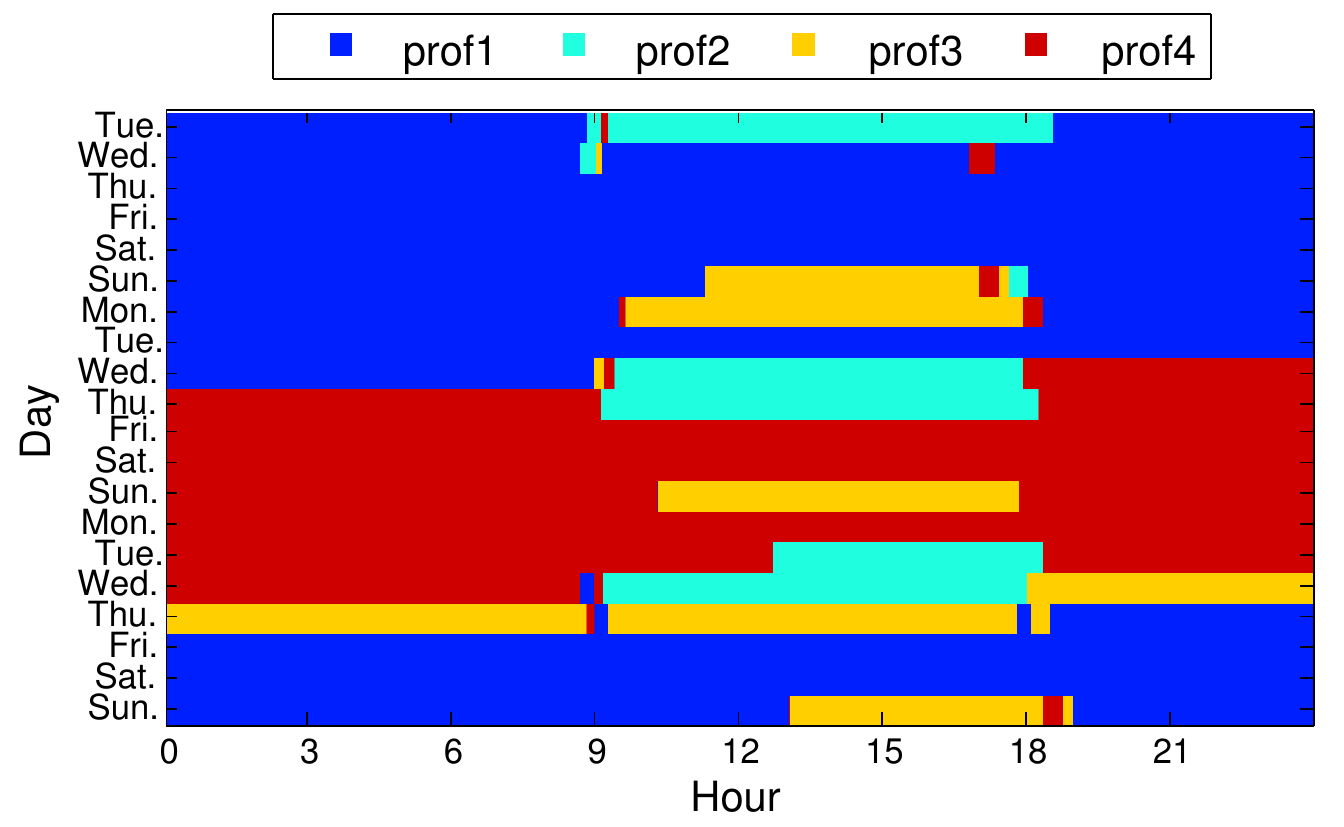}
    \caption{\small Feature-masking mitigation \emph{(p=0.4)}}
    \label{fig:profile_timeline_miti_maskf}
  \end{subfigure}
\caption{Profile timeline of user \#2 after \sysname applies the mitigation techniques\vspace{-0.0cm}}
\label{fig:profile_timeline_mitigation}
\end{figure*}

% Figure generation tip:
% 1. execute <code_clustering>/plot_accu_im.m
% 2. figures can be found in <figures>/accu_im/
\begin{figure*}[!h]
\centering
  \begin{subfigure}[b]{.32\textwidth}
    \centering
    \includegraphics[width=1\textwidth]{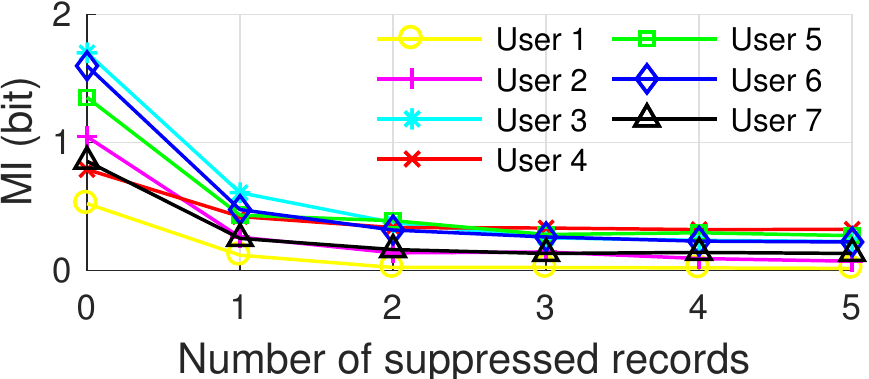}
    \caption{\small MI by suppression}
    \label{fig:im_suppress}
  \end{subfigure}
  \begin{subfigure}[b]{.32\textwidth}
    \centering
    \includegraphics[width=1\textwidth]{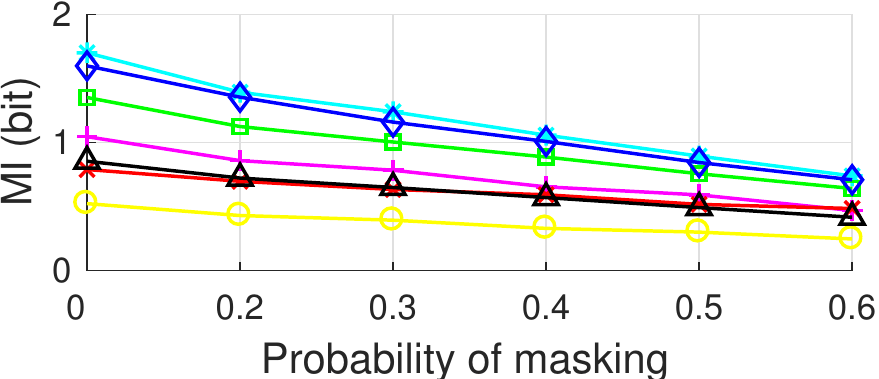}
    \caption{\small MI by row-masking}
    \label{fig:im_maskrow}
  \end{subfigure}
  \begin{subfigure}[b]{.32\textwidth}
    \centering
    \includegraphics[width=1\textwidth]{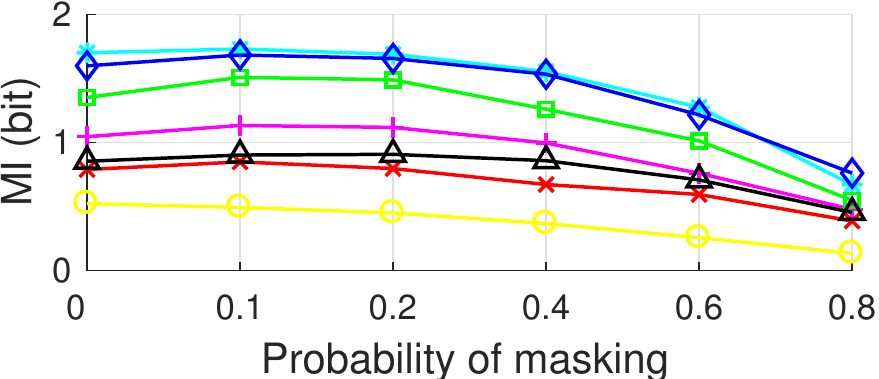}
    \caption{\small MI by feature-masking}
    \label{fig:im_maskf}
  \end{subfigure}
  
  \begin{subfigure}[b]{.32\textwidth}
    \centering
    \includegraphics[width=1\textwidth]{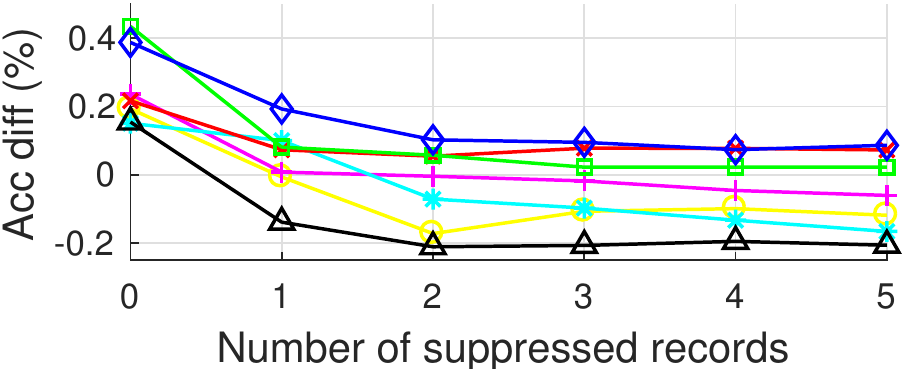}
    \caption{\small Accuracy by suppression}
    \label{fig:accu_suppress}
  \end{subfigure}
  \begin{subfigure}[b]{.32\textwidth}
    \centering
    \includegraphics[width=1\textwidth]{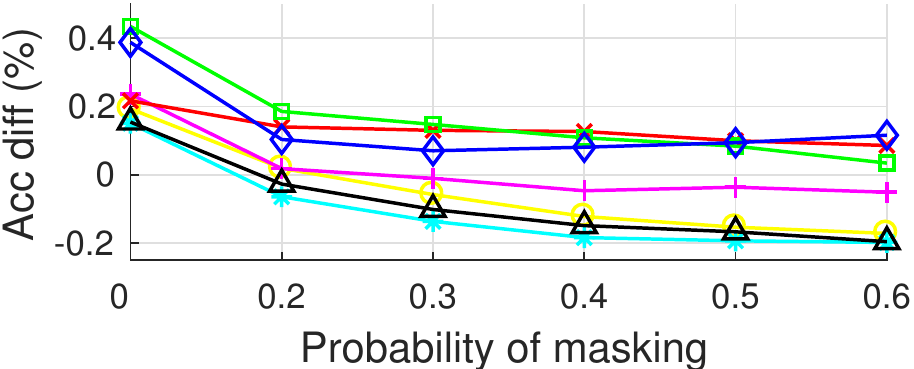}
    \caption{\small Accuracy by row-masking}
    \label{fig:accu_maskrow}
  \end{subfigure}
  \begin{subfigure}[b]{.32\textwidth}
    \centering
    \includegraphics[width=1\textwidth]{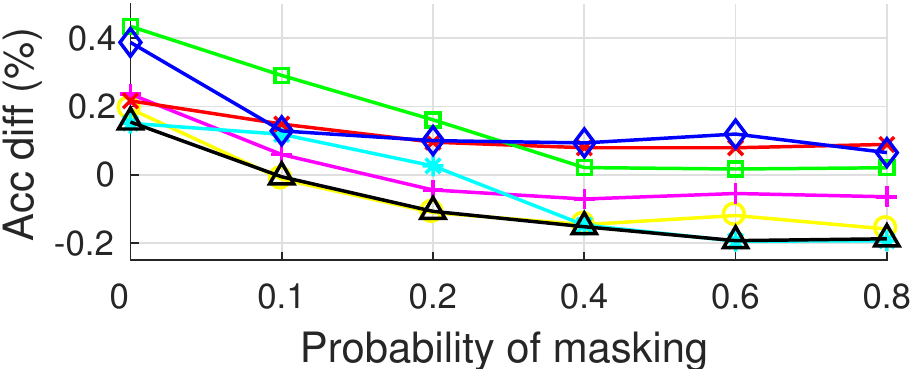}
    \caption{\small Accuracy by feature-masking}
    \label{fig:accu_maskf}
  \end{subfigure}
  \caption{Mutual information (MI) and clustering accuracy difference (Acc diff)---with respect to the baseline accuracy---after applying different mitigation methods. When mitigation magnitude increases, both mutual information and accuracy decrease}
\label{fig:mitigation}
\end{figure*}
Figure~\ref{fig:fpfn} reports the false positive and false negative rates obtained from 8000 data points collected from our user studies versus different \emph{alarm thresholds}. %For the extreme case, when the alarm threshold is set to zero, any application that possesses any \emph{slight} information is marked to be malicious by the information-based detector.% This leads to detecting any \spyware (i.e., false negative rate = 0) on the cost of an excessive false positive rate (where all benign applications are marked malicious as well).
 As the alarm threshold increases, the information-based detection becomes less aggressive leading to a significant decrease in the false positive rates while sacrificing the ability to detect malicious apps (reflected by the increase in the false negative rates). The results reported in Figure~\ref{fig:fpfn} suggest that an alarm threshold of 0.65 leads to a compromise between false positive and false negative rates (false positive rate = 0.1 and false negative rate = 0.15).
\vspace{-0.15cm}
\subsubsection{\textbf{Detection of \spyware in the Market}}
Next, we ran the proposed information-based detection algorithm (with alarm threshold set to 0.65) against the 45 real applications from the market used in Experiment 4. Recall that Experiment 4 asserts most apps can identify the user context with accuracy more than 80\% which in turn indicates that each app possesses high amount of  information. Running the proposed information-based detection algorithm against these applications results into suspicion scores 
that are greater than the alarm threshold for \emph{all} the 45 apps. Therefore, \sysname marks \emph{all} these apps as malicious reflecting the fact that these apps possess enough information to leak sensitive details about the user behavior. For space limits, we report here the suspicion score of the extreme cases (the top two and bottom two apps in terms of accuracy in Experiment 4) for the \spyware that monitors changes based on location as follows: 360 Security Antivirus (0.974), Clean Master (0.971), pinterest (0.722), and Madden NFL Mobile (0.784). Similar suspicion scores are obtained when \spyware  is monitoring changes based on calendar events.
\vspace{-0.2cm}
\subsection{Experiment 6: Performance of Mitigation Algorithms}\label{sec:evalmitigation}

To evaluate the performance of the proposed mitigation technique, we applied the proposed mitigation methods on \spyware Edge and \spyware Cloud. 
%\vspace{-0.6cm}
\subsubsection{\textbf{Mitigation of \spyware Edge}}\label{sec:mitigatespyconedge}
In Figure~\ref{fig:profile_timeline_mitigation}, we plot the profile timeline of user 2 (the profile of the same user was analyzed and discussed previously in Section~\ref{sec:spyconedge}) after applying different mitigation techniques. Comparing the results in Figure~\ref{fig:profile_timeline_selected} and Figure~\ref{fig:profile_timeline_mitigation}, we notice the distortion in the patterns of user profile compared to the case when no mitigation is applied. To better judge the effect of the proposed mitigation algorithms, we plot in Figure~\ref{fig:mitigation} the accuracy of the developed clustering algorithm when different mitigation techniques are applied along with the corresponding mutual information.  These results show how the \spyware accuracy (measured as the accuracy increase above the baseline accuracy) drops in general to the baseline accuracy. 
%(as shown in the accuracy difference in Figure~\ref{fig:mitigation}), and for some users it even drops to less than the baseline accuracy. 
 Moreover, and as expected from the theoretical underpinnings of mutual information, whenever mutual information decreases---as a consequence of the mitigation parameter  (number of records $k$ in suppression case and the probability of masking $p$ in the row-masking and feature masking case)--- the accuracy of the developed \spyware must decrease (on average).
%Generally speaking, for all mitigation methods, when stronger mitigation magnitude is applied, both mutual information and accuracy decrease, which implies that mutual information can serve as an indicator of how well a \spyware can reveal the underlying contexts. 
We also observe that the suppression method leads to a sharp decrease in the mutual information with a fast saturation whenever $k > 2$ as shown in  %the number of records suppressed is more than two records 
Figure~\ref{fig:im_suppress}. In contrast, the other two proposed mask mitigation methods (row-masking and feature-masking) lead to a gradual decrease in the mutual information as the probability of masking $p$ increases (Figure~\ref{fig:im_maskrow}, ~\ref{fig:im_maskf}). This gradual degradation provides more flexibility for \sysname to adjust to a desired mutual information. % from $XXXXX\%$ before mitigation to around $XXXXX\%$  after mitigation. }
%accuracy of the malicious \spyware---to accurately identify user profile---\r{drops from $90\%$ before mitigation to around $43\%$} after mitigation.

%Figure~\ref{fig:mitigation} shows that suppression method lead to a sharp decrease in the mutual information with a fast saturation whenever the number of records suppressed is more than two records. In contrast, the two proposed mask mitigation methods lead to a gradual decrease in the mutual information as the probability of masking increases. This gradual degradation provides more flexibility for \sysname to adjust to a desired mutual information. Row-masking even shows a close inverse-proportional relation between mask probability and mutual information.

Finally, though we expect mutual information should monotonically decrease when the mitigation is stronger (higher masking probability), \ref{fig:im_maskf} shows that mutual information goes higher when masking probability $p=0.1$. One possible explanation is that noisy settings are cleared with this configuration, causing higher correlation between mitigated data and user profile. However, the mutual information decreases sufficiently when larger masking probability is applied (i.e., $p>=0.4$).
\begin{figure*}[!t]
\centering
  \begin{subfigure}[b]{.33\textwidth}
    \centering
    \includegraphics[width=1\textwidth]{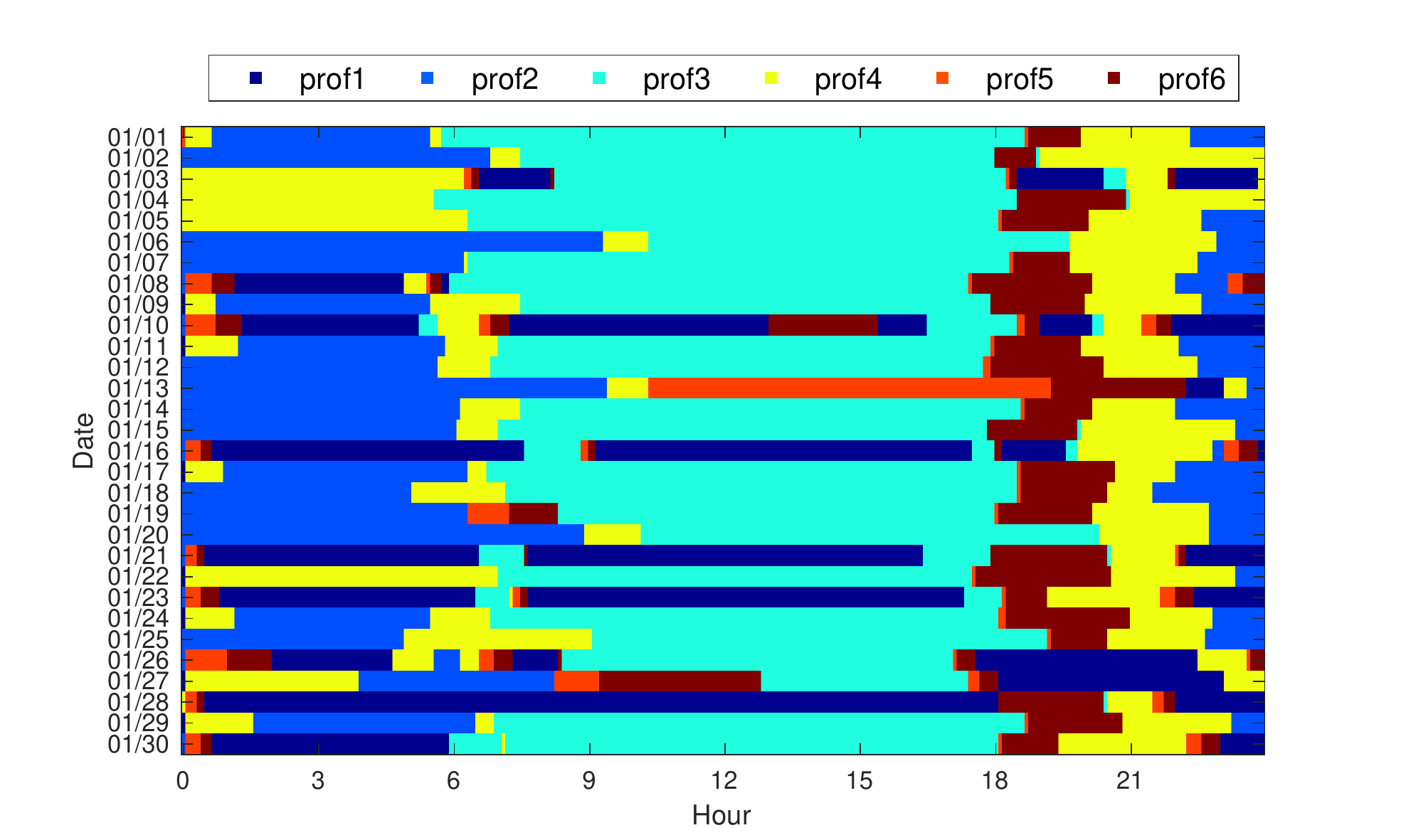}
    \caption{\small Profiles detection after masking mitigation \emph{(p=0.4)}}
    \label{fig:activity_mitigation_p0.4}
  \end{subfigure}
  \begin{subfigure}[b]{.33\textwidth}
    \centering
    \includegraphics[width=1\textwidth]{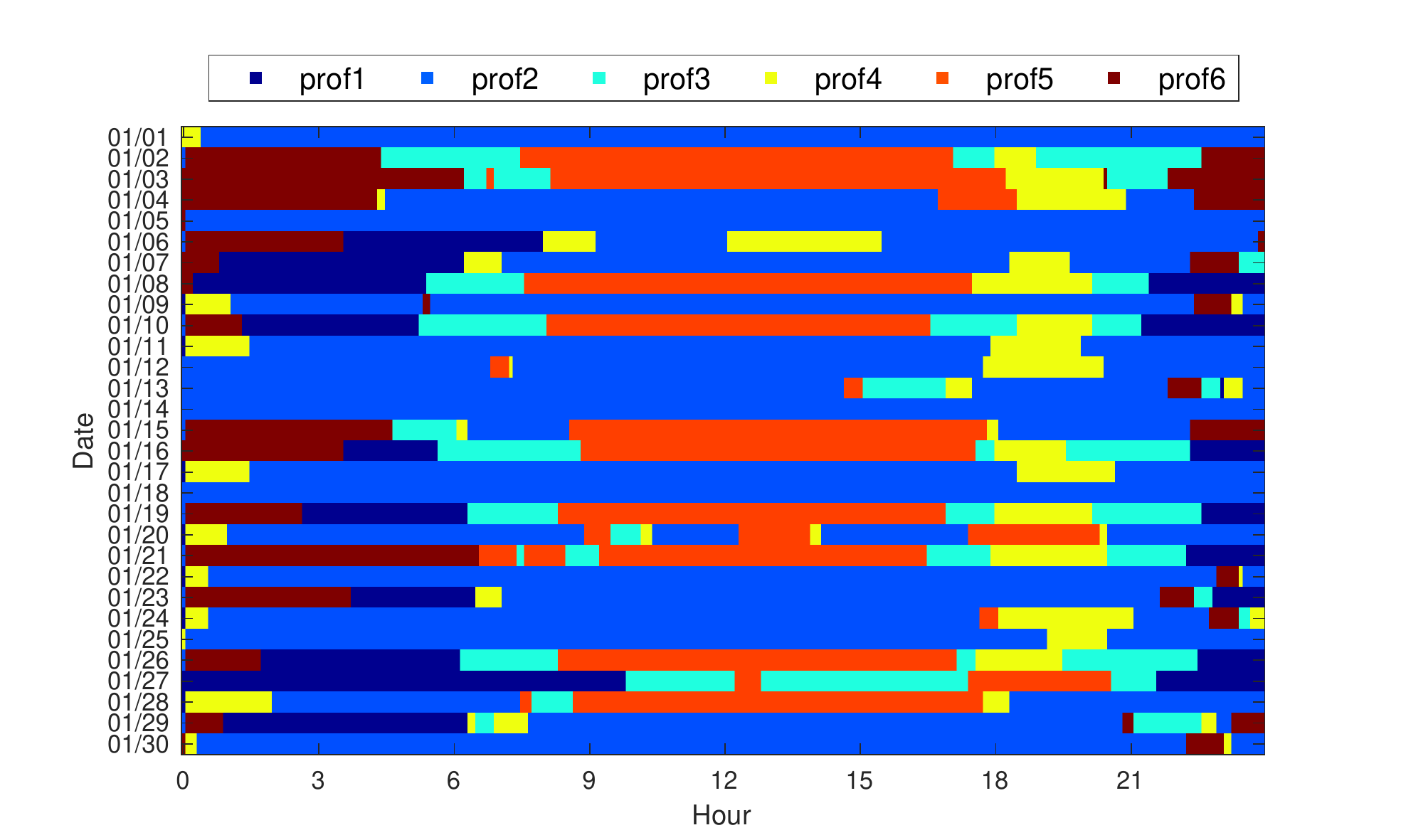}
    \caption{\small Profiles detection after masking mitigation \emph{(p=0.8)}}
    \label{fig:activity_mitigation_p0.8}
  \end{subfigure}
  \begin{subfigure}[b]{.33\textwidth}
    \centering
    \includegraphics[width=1\textwidth]{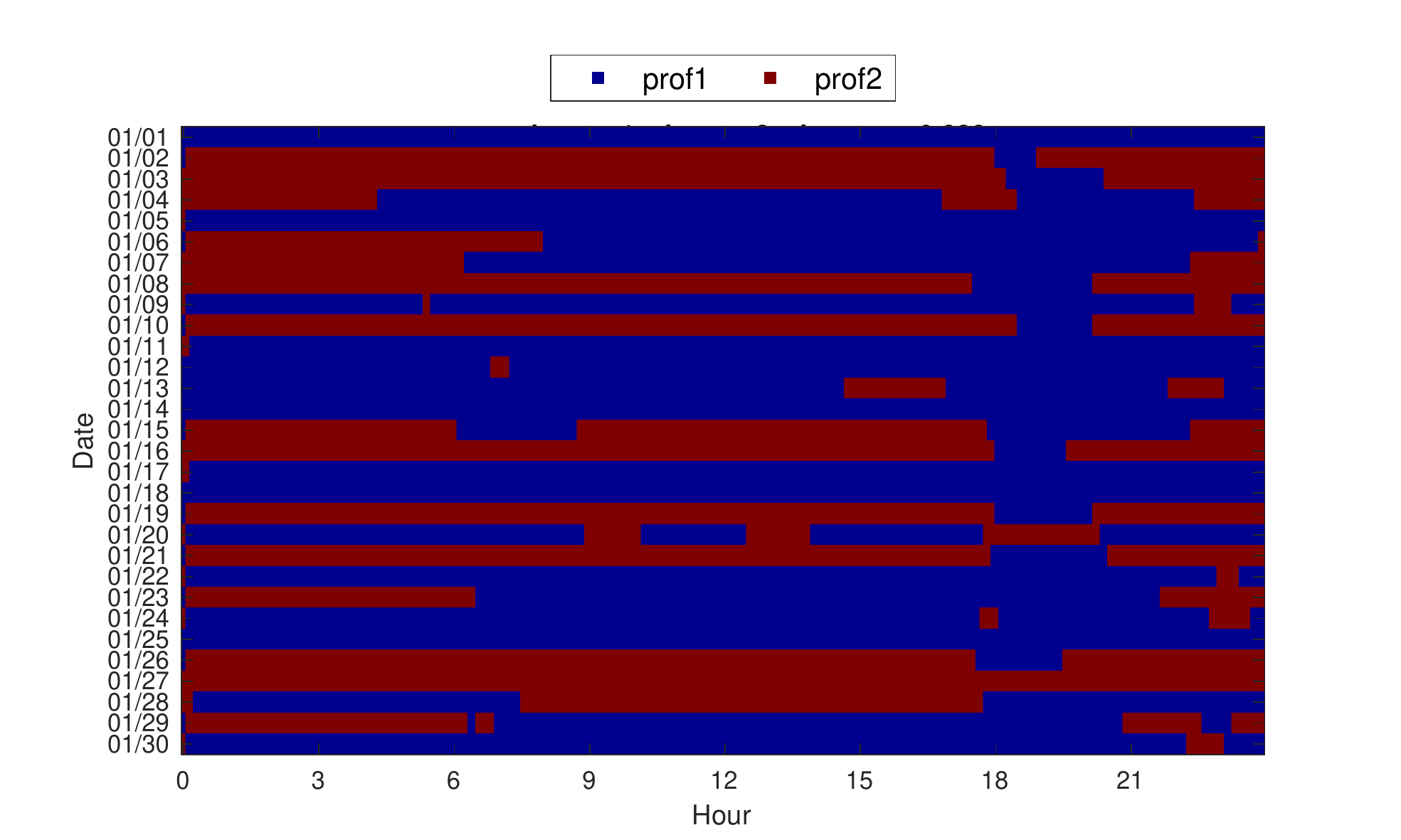}
    \caption{\small Occupancy detection after masking mitigation \emph{(p=0.8)}}
    \label{fig:occupancy_mitigation_p0.8}
  \end{subfigure}
\caption{Activity and occupancy timeline of human \#1 after \sysname applies the mitigation techniques \vspace{-2mm}}
\label{fig:hvac_timeline_mitigation}
\end{figure*}

\subsubsection{\textbf{Mitigation of \spyware Cloud}}
\begin{figure}[!h]
\centering
  \begin{subfigure}[b]{.49\columnwidth}
    \centering
    \includegraphics[width=0.8\textwidth]{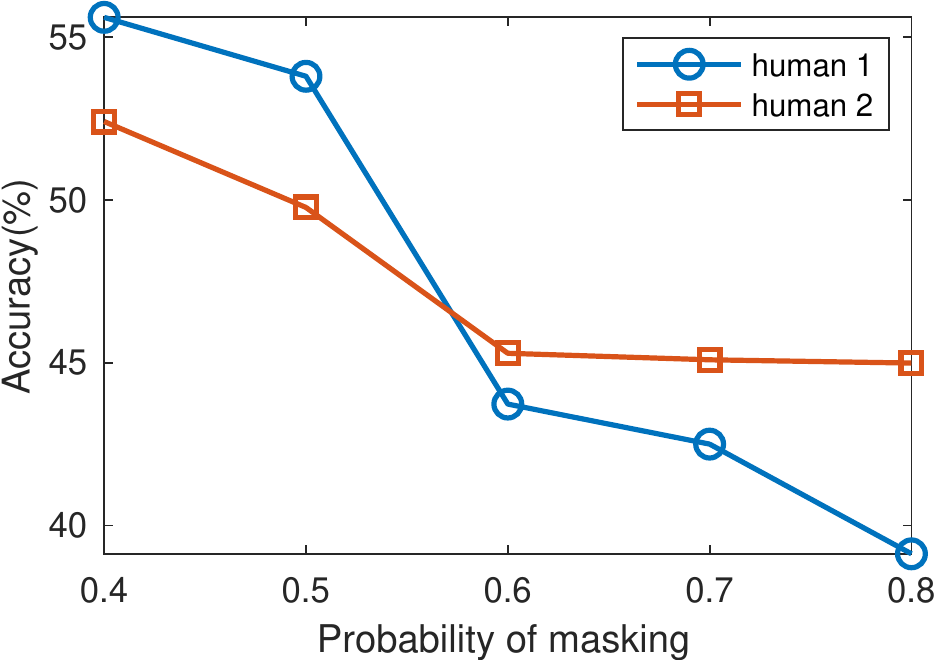}
    \caption{\small Accuracy of activity detection after mitigation}
    \label{fig:mitigation_acc_activity}
  \end{subfigure}
  \begin{subfigure}[b]{.49\columnwidth}
    \centering
    \includegraphics[width=0.8\textwidth]{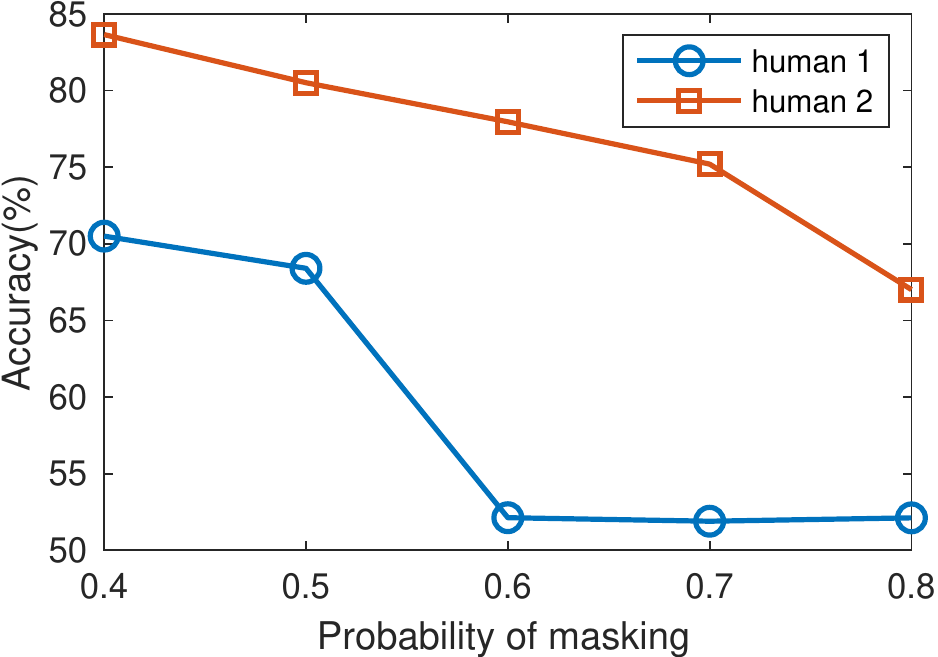}
    \caption{\small Accuracy of occupancy detection after mitigation}
    \label{fig:mitigation_acc_occupancy}
  \end{subfigure}
  \caption{Clustering accuracy after applying mitigation methods on \spyware Cloud. When mitigation magnitude increases, the accuracy decreases\label{fig:mitigation_hvac}}
\end{figure}

Using the same mitigation procedure as \spyware Edge (Section~\ref{sec:mitigatespyconedge}), we plot the mitigation results of human \#1 in Figure~\ref{fig:hvac_timeline_mitigation}. We applied row masking and feature masking simultaneously with different probabilities. We observe that the performance of the \spyware Cloud in detecting the user profile is adversely affected by increasing the probability of masking. Similarly, as shown in Figure~\ref{fig:occupancy_mitigation_p0.8}, the \spyware ability to detect home occupancy degraded severely as a consequence of applying the mitigation. The decrease in accuracy for both human 1 and human 2 across different probabilities of mitigation is shown in Figure~\ref{fig:mitigation_hvac}.

\subsubsection{\textbf{Effect of Mitigation on Human Thermal Comfort}}
\begin{figure*}[!h]
\centering
  \begin{subfigure}[b]{.33\textwidth}
    \centering
    \includegraphics[width=1\textwidth, height=2cm]{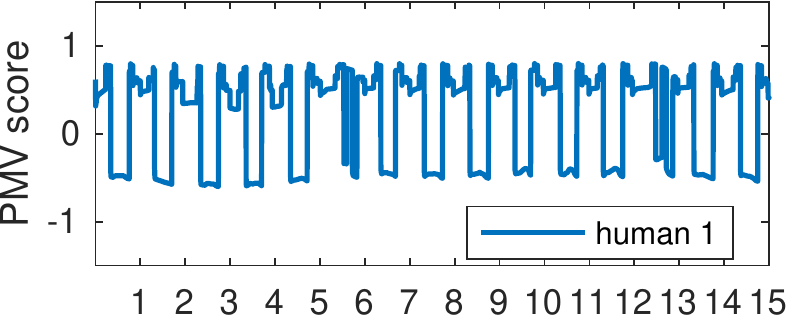}
    \caption{\small Before mitigation}
    \label{fig:pmv_normal}
  \end{subfigure}
  \begin{subfigure}[b]{.33\textwidth}
    \centering
    \includegraphics[width=1\textwidth, height=2cm]{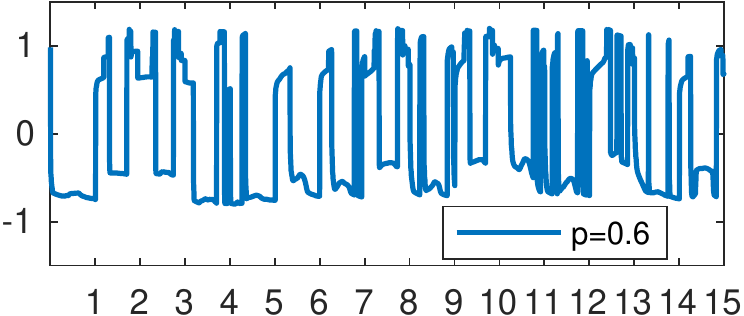}
    \caption{\small After mitigation by masking($p=0.6$)}
    \label{fig:pmv_mitigate_06}
  \end{subfigure}
    \begin{subfigure}[b]{.33\textwidth}
    \centering
    \includegraphics[width=1\textwidth, height=2cm]{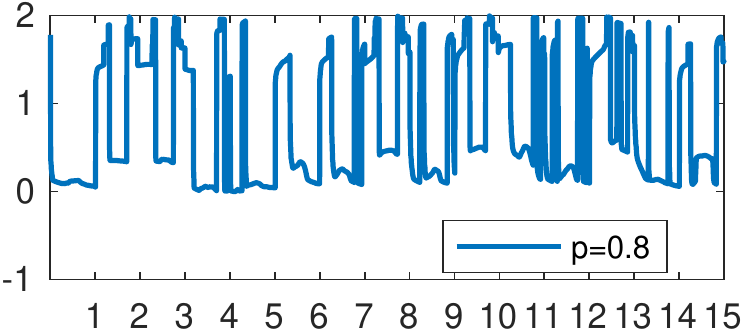}
    \caption{\small After mitigation by masking($p=0.8$)}
    \label{fig:pmv_mitigate_08}
  \end{subfigure}
  \caption{PMV score before and after mitigation with different masking probabilities across 15 days \label{fig:pmv} \vspace{-5mm}}
\end{figure*}

Applying row and feature masking in the context of \spyware Cloud entails changing the HVAC set-point in order to hinder the ability of \spyware  to infer the human behavior can affect the human thermal comfort. Accordingly, we compared the Prediction Mean Vote (PMV) values---as a measure for the human thermal comfort~\cite{fanger1970thermal}---before and after mitigation in Figure~\ref{fig:pmv}. The PMV score ranges from $-3$ to $3$ which is the range of thermal sensation from very cold ($-3$) to very hot ($3$). According to ISO standard ASHRAE 55~\cite{handbook2009american}, a PMV in the range of $-0.5$ and $+0.5$ for an interior space is recommended to achieve thermal comfort. We used the Fanger model in EnergyPlus to estimate the PMV value~\cite{fanger1970thermal}. In particular, as seen in Figure~\ref{fig:pmv}, a choice of masking probability equal $0.6$ lead to user PMV in the range of $-0.8$ to $1.2$ while achieving a degradation in \spyware accuracy by $45\%$ (as shown in Figure~\ref{fig:mitigation_hvac}).

%\begin{figure*}[!h]
%\centering
%  \begin{subfigure}[b]{.15\textwidth}
%    \centering
%    \includegraphics[width=1\textwidth]{figures/smarthome/figures/PMV/PMV_normal.eps}
%    \caption{\small PMV }
%    \label{fig:pmv_normal}
%  \end{subfigure}
%  \begin{subfigure}[b]{.15\textwidth}
%    \centering
%    \includegraphics[width=1\textwidth]{figures/smarthome/figures/PMV/PMV_h1_mitigated_04.eps}
%    \caption{\small $p=0.4$}
%    \label{fig:pmv_04}
%  \end{subfigure}
%   \begin{subfigure}[b]{.15\textwidth}
%    \centering
%    \includegraphics[width=1\textwidth]{figures/smarthome/figures/PMV/PMV_h1_mitigated_05.eps}
%    \caption{\small $p=0.5$}
%    \label{fig:pmv_05}
%  \end{subfigure}
%    \begin{subfigure}[b]{.15\textwidth}
%    \centering
%    \includegraphics[width=1\textwidth]{figures/smarthome/figures/PMV/PMV_h1_mitigated_06.eps}
%    \caption{\small $p=0.6$}
%    \label{fig:pmv_06}
%  \end{subfigure}
%      \begin{subfigure}[b]{.15\textwidth}
%    \centering
%    \includegraphics[width=1\textwidth]{figures/smarthome/figures/PMV/PMV_h1_mitigated_07.eps}
%    \caption{\small $p=0.7$}
%    \label{fig:pmv_07}
%  \end{subfigure}
%        \begin{subfigure}[b]{.15\textwidth}
%    \centering
%    \includegraphics[width=1\textwidth]{figures/smarthome/figures/PMV/PMV_h1_mitigated_08.eps}
%    \caption{\small $p=0.8$}
%    \label{fig:pmv_08}
%  \end{subfigure}
%  \caption{PMV score before and after mitigation with different masking probabilities.\label{fig:pmv}
%\vspace{-0.4cm}  
%   %\r{redraw the figures with bigger space for the caption}
%  }
%\end{figure*}

\vspace{-0.15cm}
\section{Conclusion}\label{sec:conclusion} 
In this paper, we introduced \spyware, a new class of privacy threatening spyware that monitors the adaptation actions and the decisions made by HITL IoT in an attempt to leak sensitive information about the IoT user. We showed two concrete examples of \spyware namely \spyware Edge and \spyware Cloud. In the first example (\spyware Edge), we showed through our user study how a \spyware Edge could infer user behavior and the semantics of the user location in realtime. To exacerbate the situation, our experiments showed that this new spyware is undetectable using off-the-shelf antivirus packages and moreover many of the top 45 downloadable free Android apps have enough information to reveal sensitive user information. In the second example (\spyware Cloud), we showed a simulated example of a smart HVAC, in which the cloud service has a mounted \spyware. In our experiments, we showed how the mounted \spyware can infer the user's occupancy at home and his daily schedule (awake, sleep).

To circumvent such scenarios, we designed \sysname, a safeguard which protects authentic HITL IoT applications against leaking private information via this side-channel. \sysname employs a detection technique based on mutual information which is agnostic to the \spyware implementation details. \sysname uses three mitigation techniques. Our mitigation techniques have shown a degradation of \spyware Edge inference accuracy from $90.3\%$ to the baseline accuracy and a degradation in \spyware Cloud by $45\%$ and by only adding negligible overhead ($3\%$) on the API call performance. %Due to the lack of systematic support that targets the attacks on context-aware adaptation apps, \sysname comes to fill in this gap.
Future work includes studying the effect of collaborative \spyware. In such scenario, multiple IoT devices collect different (or partial) information from the proposed side-channel and fuse them to leak sensitive user information while maintaining small individual mutual information and hence bypass the information-based detection algorithm. %Another direction is to extend \sysname beyond context-aware applications by learning the context-actions relations at runtime without the need of a registry file that is generated at installation time. 

\vspace{-4mm}
\bibliographystyle{ACM-Reference-Format}
\bibliography{sigproc}

%%
%% If your work has an appendix, this is the place to put it.
\appendix

%#####################################################################################
%\iffalse
%
%\section{Research Methods}
%\subsection{Part One}
%Lorem ipsum dolor sit amet, consectetur adipiscing elit. Morbi
%malesuada, quam in pulvinar varius, metus nunc fermentum urna, id
%sollicitudin purus odio sit amet enim. Aliquam ullamcorper eu ipsum
%vel mollis. Curabitur quis dictum nisl. Phasellus vel semper risus, et
%lacinia dolor. Integer ultricies commodo sem nec semper.
%
%\subsection{Part Two}
%Etiam commodo feugiat nisl pulvinar pellentesque. Etiam auctor sodales
%ligula, non varius nibh pulvinar semper. Suspendisse nec lectus non
%ipsum convallis congue hendrerit vitae sapien. Donec at laoreet
%eros. Vivamus non purus placerat, scelerisque diam eu, cursus
%ante. Etiam aliquam tortor auctor efficitur mattis.
%
%\section{Online Resources}
%Nam id fermentum dui. Suspendisse sagittis tortor a nulla mollis, in
%pulvinar ex pretium. Sed interdum orci quis metus euismod, et sagittis
%enim maximus. Vestibulum gravida massa ut felis suscipit
%congue. Quisque mattis elit a risus ultrices commodo venenatis eget
%dui. Etiam sagittis eleifend elementum.
%Nam interdum magna at lectus dignissim, ac dignissim lorem
%rhoncus. Maecenas eu arcu ac neque placerat aliquam. Nunc pulvinar
%massa et mattis lacinia.
%
%
%\fi 
% ####################################################################################

\end{document}